\documentclass[10pt,letterpaper]{article}

\usepackage[utf8]{inputenc}

\usepackage{natbib}

\usepackage{nameref,hyperref}
\usepackage{longtable}
\usepackage[right]{lineno}

\usepackage{microtype}
\DisableLigatures[f]{encoding = *, family = * }

\raggedright
\usepackage{changepage}

\usepackage[aboveskip=1pt,labelfont=bf,labelsep=period,singlelinecheck=off]{caption}

\makeatletter
\renewcommand{\@biblabel}[1]{\quad#1.}
\makeatother

\usepackage{lastpage,fancyhdr,graphicx}
\usepackage{epstopdf}
\fancyheadoffset[L]{2.25in}
\fancyfootoffset[L]{2.25in}

\usepackage{color}

\definecolor{Gray}{gray}{.25}

\usepackage{graphicx}

\usepackage{sidecap}

\usepackage{wrapfig}
\usepackage[pscoord]{eso-pic}
\usepackage[fulladjust]{marginnote}
\reversemarginpar

\begin{document}
\vspace*{0.35in}

\begin{flushleft}
{\Large
\textbf\newline{First Solar Radio Burst Observations by the Mexican Array Radio Telescope (MEXART) at 140 MHz}
}
\newline
\\

E. Huipe-Domratcheva\textsuperscript{1},
V. De la Luz\textsuperscript{2},
G. A. Casillas-Perez\textsuperscript{3},
J. C. Mejia-Ambriz\textsuperscript{3},
E. Perez-Leon\textsuperscript{5},
J. A. Gonzalez-Esparza\textsuperscript{6},
C. Monstein\textsuperscript{7}
W. Reeve\textsuperscript{8}
\\
\bigskip
\bf{1} Posgrado de Ciencias de la Tierra,Escuela Nacional de Estudios Superiores Unidad Morelia, Universidad Nacional Autonoma de Mexico, Morelia, 58190, Mexico
\\
\bf{2} Escuela Nacional de Estudios Superiores Unidad Morelia, Universidad Nacional Autonoma de Mexico, Morelia, 58190, Mexico
\\
\bf{3} Instituto de Geofisica, Universidad Nacional Autonoma de Mexico, Ciudad Universitaria, C.P. 04510 Ciudad de Mexico, Mexico
\\
\bf{4} Facultad de Ciencias Fisico Matematicas, Universidad Autonoma de Nuevo Leon, San Nicolas de los Garza, 66455, Mexico
\\
\bf{5} Instituto de Geofisica Unidad Michoacan, Universidad Nacional Autonoma de Mexico, Morelia, 58190, Mexico
\\
\bf{6} Istituto Ricerche Solari (IRSOL), Universita della Svizzera italiana (USI), CH-6605 Locarno-Monti, Switzerland
\\
\bf{7} Reeve Observatory, Anchorage, Alaska, USA
\\
\bigskip
* vdelaluz@enesmorelia.unam.mx

\end{flushleft}

\section*{Abstract}
The National Laboratory of Space Weather in Mexico (Laboratorio Nacional de Clima Espacial: LANCE) coordinates instrumentation for
monitoring the space-weather impact over Mexico. Two of these instruments are the \textit{Mexican Array Radio Telescope} (MEXART) and \textit{Compound Astronomical Low frequency Low cost Instrument for Spectroscopy and Transportable Observatory} (CALLISTO) station from the e-CALLISTO network (CALLMEX). Both instruments are located at the same facility (Coeneo Michoacan, Mexico) and share a spectral band centered at 140\,MHz. 
In this work we show the capabilities of the e-CALLISTO network as support to 
identify  a solar radio  burst 
in the signal of the MEXART radiotelescope. 
We identify 75 solar radio bursts in the MEXART signal: five events of Type II and 70 of Type III 
 between 
September 2015 and May 2019.
The analysis of solar radio bursts in the MEXART signal provide us valuable information about 
the development of the radio event due their high sensitivity, time resolution, and isotropic 
response.  For the case of Type III solar radio events, we identify three characteristic phases
in the dynamical evolution of the signal at 140\,MHz: a pre-phase, a main peak, a decay 
phase, and a post-event phase. A Morlet wave transform was done of MEXART signals in the Type 
III solar radio busts; in their spectra it was identified a pine tree structure preceding the 
main event in the time series. These characteristics are not observable in the data from the 
e-Callisto network.
\section{Introduction}
     \label{Introduction} 
The first detection of solar radio bursts (SRB) was 
recorded as radio noise by radar antennas during
WW\,II \cite{Southworth1982,hey1946solar,pawsey1946radio}. These 
signals in the electromagnetic spectrum  originate in the solar corona and their main characteristics include
a spontaneous 
and intense
emission above the radio background over a wide range of 
frequencies depending on the SRB type \cite{raulin2005solar}. 

The flux of SRB is an indirect source of information 
for the
physical conditions of the environment 
where the radiation originates \cite{Warmuth2005}.
There are many types of SRB; even when some events are difficult to classify, the majority of events fit into five classes of burst (I\,--\,V). This classification is based on their evolution and shape 
in the dynamic spectrum \cite{pawsey1946observation,payne1947relative}, the most frequently studied in relation to space weather are the SRB Type II and III \cite{white2007solar}.

The most common instrumentation focused on the systematic record of radio flux from SRB includes radiometers, radiospectrographs, and radioheliographs \cite{payne1949bursts,bougeret1995waves,nindos2008radio}. However, for the case of intense SRB events,interference with terrestrial radio communications and sensitive radio telescopes is often considered as unknown noise \cite{appleton1945departure}. In this sense, the detection
and characterization of SRB
in real time allows us to increase our resilience in radio communications 
\cite{Warmuth2005} and provides valuable information in astrophysics radio facilities to avoid spurious signals superimposed on astronomical sources \cite{monstein2011catalog}.

The instrumentation of the National Laboratory of Space Weather (Laboratorio Nacional de Clima Espacial: LANCE) includes 
MEXART is a composite antenna of 4096 full-wave dipoles (2.14\,m each) covering near 9600\,m$^2$. The instrument operate at a center frequency of 139.65\,MHz (hereafter 140\,MHz) and 1.5\,MHz bandwidth. The antenna is composed of 64 parallel lines aligned in the East--West direction with 64 dipoles each. are a band at140\,MHz.

For many years, unknown signals recorded by the MEXART radio telescope were avoided in the main pipeline analysis. These data were classified as generic noise on the signal. However, since the installation of the CALLISTO station in 2014 we have the possibility of identifying SRB recorded with CALLISTO and comparing them against the signal of the MEXART radiotelescope. 
%
We verify for the first time that MEXART is detecting SRB. The temporal resolution and sensitivity of MEXART allow us to characterize the morphology of SRB at140\,MHz and provide valuable information about their dynamic evolution.

This work is organized as follows: Section \ref{instrumentation} describes the instrumentation and their technical configuration; in Section \ref{radiointerferences} we present an analysis of the characteristics of the electromagnetic spectrum on site between 45\,MHz and 225\,MHz; in Section \ref{observations} we present the methodology used to identify SRB in the MEXART signal and a summary of the different types of SRB observed by both instruments; in Section \ref{database} we introduce the SRB recorded from September 2015 to May 2019 for both instruments; in Section \ref{dataanalysis} we present a statistical and wavelet analysis of the SRB recorded by MEXART; and finally in Section \ref{conclusions} we present the conclusions and discussion. 
\section{Instrumentation}\label{instrumentation}
%
%
%
 \begin{figure}[p]    
   \centering{
   \includegraphics[width=1\textwidth,clip=]{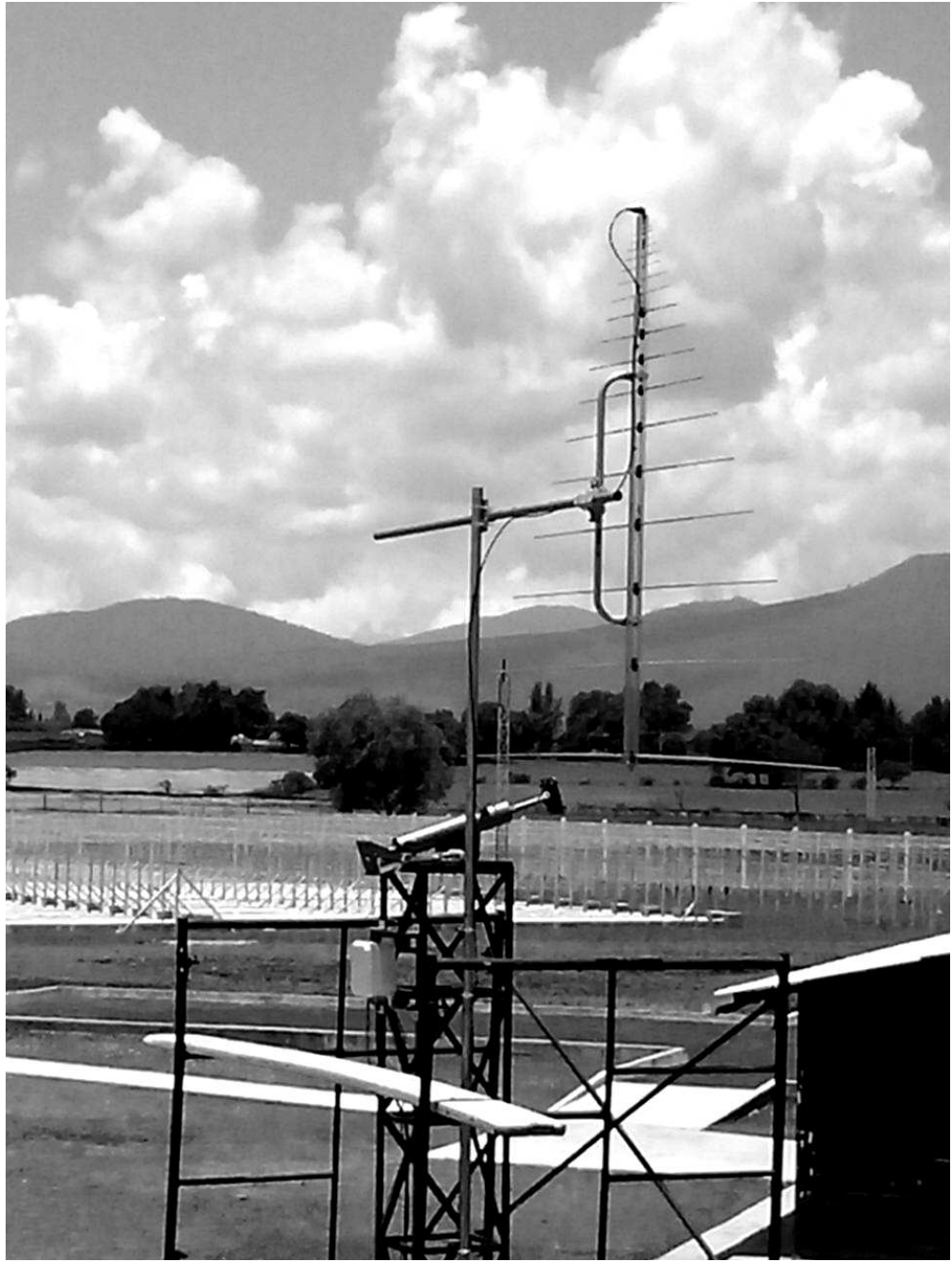}
              }
   \caption{LANCE facilities in Coeneo, Michoacan, Mexico. In the foreground the CALLISTO station CALLMEX and in the background the MEXART radio telescope with 2.14\,m dipoles covering near 9600\,m$^2$.}
    \label{callisto-picture.eps}
  \end{figure}
The main instrumentation facilities of LANCE are located in Coeneo in the state of Michoacan, Mexico $(19^{\circ} 48^{\prime} 49^{\prime\prime} N, 101^{\circ} 41^{\prime} 39^{\prime\prime} W)$ at an altitude of $1964$ masl. This location presents low radio interference levels \cite{esparza2004mexart}. Since 2005, LANCE operates the instrumentation installed in Coeneo \cite{Delaluz2018} including the MEXART radio telescope and the CALLISTO radiospectrograph, which operate in the VHF range (30\,--\,300\,MHz) (see Figure \ref{callisto-picture.eps}). 
%

MEXART is a transit instrument principally dedicated to monitoring radio sources that exhibit IPS, these are compact sources ($<1$ arcsec), typically active galactic nuclei.
The sources are recorded during its transit at the local meridian by $1^{\circ}$  beams in the East--West direction, thus compact sources are identified as an increment in the intensity of the signal during nearly four minutes. The spectral analyses and level of IPS is used to remote sense solar-wind properties along the line of sight, such as density and speed \cite{Mejia-Ambriz2010}. MEXART is continuously monitoring the radio sky, it allows detecting noise and signals the Sun during the day.


In contrast, CALLISTO scans the electromagnetic spectrum isotropically in a wide bandwidth from 45\,MHz to 225\,MHz to detect sudden emission generated by eruptive events in the solar atmosphere \cite{Benz2005}. 

The characteristics shared by both 
instruments (frequency range of observations and localization), even when the instruments have different scientific focus, allow us to identify and validate signals of SRB with high temporal resolution at 140\,MHz. 

In the following subsections we describe the technical characteristics and configurations of both instruments used in this work.

\subsection{MEXART Radiotelescope}
MEXART is a composite antenna of 4096 full-wave dipoles (2.14\,m each) covering near 9600\,m$^2$. The instrument operate at a center frequency of 139.65\,MHz (hereafter 140\,MHz) and 1.5\,MHz bandwidth. The antenna is composed of 64 parallel lines aligned in the East--West direction with 64 dipoles each.

MEXART is designed to indirectly measure the solar-wind speed by using IPS. The  telescope operates in transit mode, i.e. detecting radio sources passing through the local meridian. Due to the rotation of the Earth, its fan-beam antenna power pattern scans the sky once each sidereal day \cite{Mejia-Ambriz2010}. 
A beam-former system is produced by a Butler Matrix to produce 16 fixed beams at different declinations in the local meridian.
Computational methodologies are used to store and handle the instrument's real-time data \cite{Casillas2010}. The data are sampled with a cadence of $\approx20$ milliseconds. A computer displays the signal in real time. Data copies are stored each 24 hours as a single file, one in a local server and other one in our local repository \cite{gonzalez2017mexican}. 


\subsection{CALLISTO Station at MEXART Facility (CALLMEX)}
The CALLISTO-MEXART station (CALLMEX) is a radiospectrograph that belongs to the e-CALLISTO network, responsible for
monitoring the Sun continuously along with other 164 stations spread  globally \cite{ecallisto}. The e-CALLISTO network 
is focused on: the detection of solar radio bursts, study of radiointerference, and other educative purposes \cite{Benz2005}.

The aim of the network is to monitor solar activity events, particularly CMEs (Coronal Mass Ejections) and flares, through their radio emissions. The different CALLISTO stations that conform the network have different facilities such as type of antennas like Yagi or LPDA or even dish antennas. Some receivers have modifications in their gain, number of channels, etc. The data produced by these instruments in general cover the VHF band  and in lower proportion in HF (3\,--\,30\,MHz) and UHF (300\,--\,3000\,MHz).


 The CALLMEX station has a radio-frequency receiver, coupled to the sky by a log-periodic dipole array (LPDA),  which allows the detection of a wide bandwidth from 50\,MHz to 1.3\,GHz with a gain between 10\,--\,12\,dBi. 
 The received signals travel through a coaxial cable to a low noise amplifier (TMA-1) at 20\,dB.
 The TMA-1 is also equipped with a rod arrester and a peak suppressor for other sudden and strong signals that could damage the equipment due to high voltage peaks. The signal travels from the amplifier to the main receiver. Finally, in the control room the signal is processed.



The CALLISTO receiver allows us to process signals between 45 and 870\,MHz. The first configuration of CALLMEX was established between 45\,--\,354\,MHz. In February 2016 after first tests, the configuration changed to 45\,--\,225\,MHz. 

The station records data from 13:15 UTC to 00:30 UTC (of the 
UTC next day) which corresponds to local daylight. 


In the following section we present the main characteristics of the  electromagnetic spectrum observed by MEXART and CALLMEX. 

\section{Electromagnetic Spectrum between 45 MHz and 225 MHz at Site}
      \label{radiointerferences}
We characterize the electromagnetic spectrum between 45\,MHz and 225\,MHz 
using  
CALLMEX. 
CALLISTO detects 
local radio emissions generated by communications, lightning, or noise from the electric power network, in the E--W direction. The important feature of CALLISTO is that it observes in a wide spectral window, which allows the detection of signals and displays them in dynamic spectra. 

\subsection{Radio-Broadcast}
There are emissions that affect the analysis of solar radio events; one of the main interference sources is 
telecommunications. To identify the origin of such it is necessary to request and inspect  the logs of the Federal
Communications Office (IFT by its acronym in Spanish), which regulates the telecommunications of civil organisms, the government,
private sector, etc. In this way, the IFT  assigns operating bands that can be monitored by CALLMEX. According to the National Table of Frequency Allocations \cite{leyfederal}, the main operations are:
\begin{itemize}
    \item Amateur radio, transmitting or receiving signals (Tx-Rx)
    \item Radioastronomy (Rx)
    \item Radionavigation for areonautics (Rx-Tx)
    \item Broadcasting (Tx)
    \item Space research (satellites, tethorology, radiolocation) (Rx-Tx)
    \item Rescue and emergency calls (Rx-Tx)
    \item Private use (Rx-Tx)
    \item Commercial television and radio (Rx-Tx)
\end{itemize}



The spectrum recorded by CALLISTO is binned in 200 channels covering the whole operating bandwidth. Two bands between 90\,--\,110\,MHz and 125\,--\,172\,MHz contain anthropogenic signals, see first spectrum of  Figure \ref{spectrums}. In the first band are located television and FM radio stations. The second band detects satellite signals that are emitted in the MEXART band as shown by \cite{carrillo2012ionospheric}. 

\subsection{Noise at MEXART Facilities}
\begin{figure}
  \centering
  \caption{Example of CALLMEX spectra in its actual bandwidth (45\,--\,225\,MHz). In the first spectrum it can be seen the local noise, predominately the FM radio between 88 and 110\,MHz. Zebra emission noise produced by local electronic RFI in the second spectrum. In the third spectrum it is shown a natural RFI source, the emission of lightnings, which can affect the studies of astronomical radio sources in CALLMEX and MEXART. The fourth spectrum shows a solar transit using one line of MEXART as antenna.}
  \begin{tabular}{cc}
    \includegraphics[width=56mm]{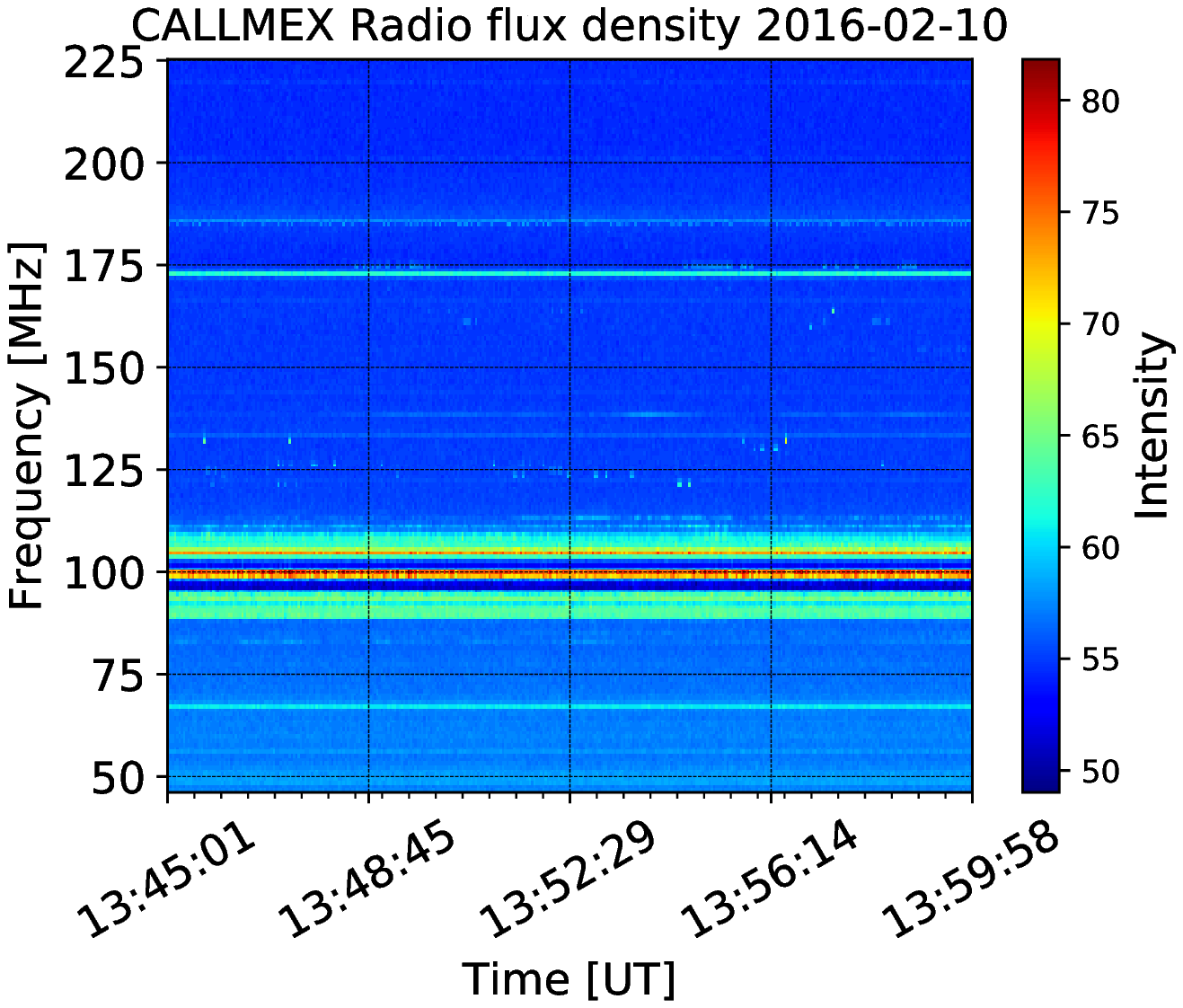}
    \includegraphics[width=56mm]{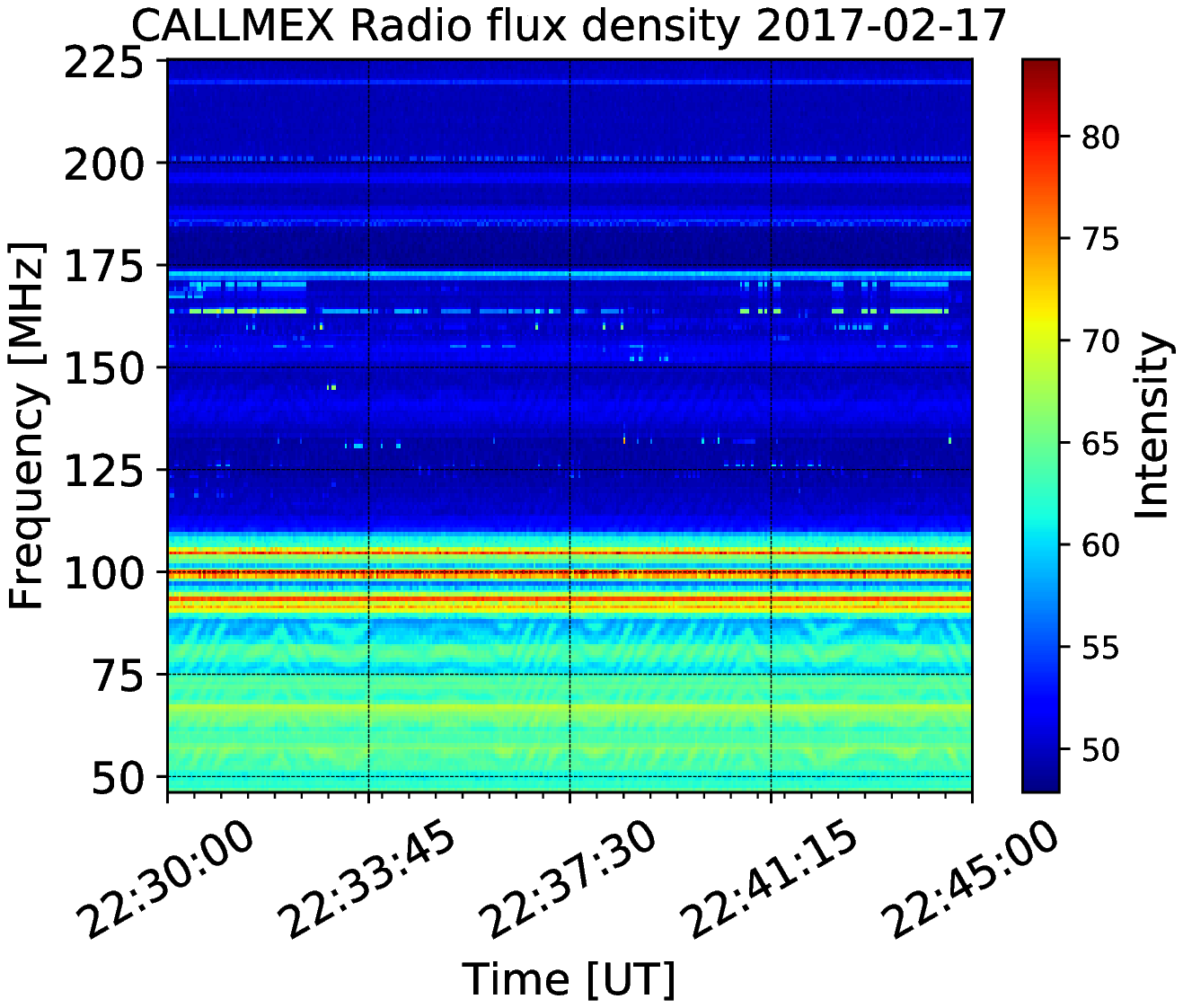}\\
     \includegraphics[width=56mm]{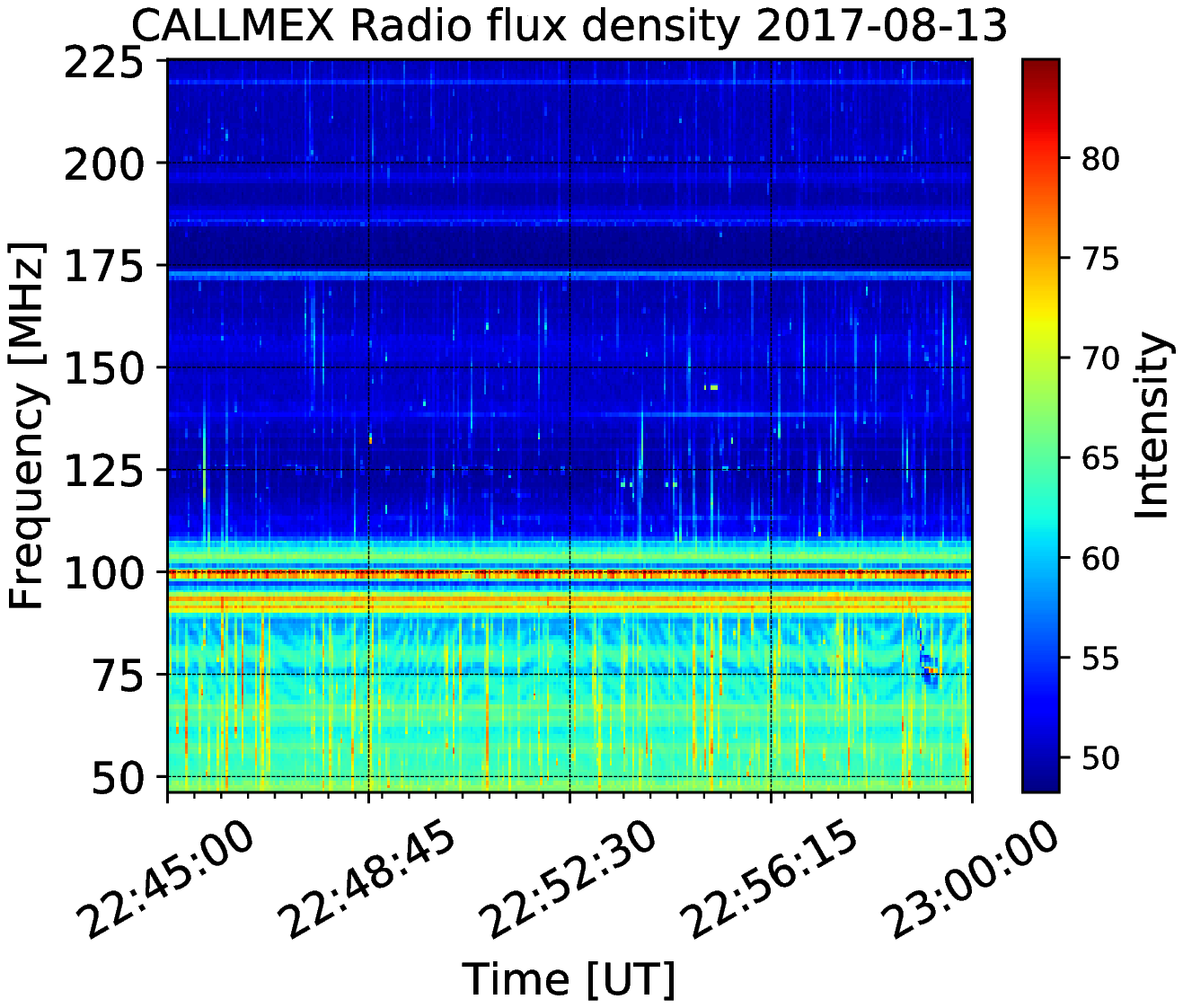}
     \includegraphics[width=56mm]{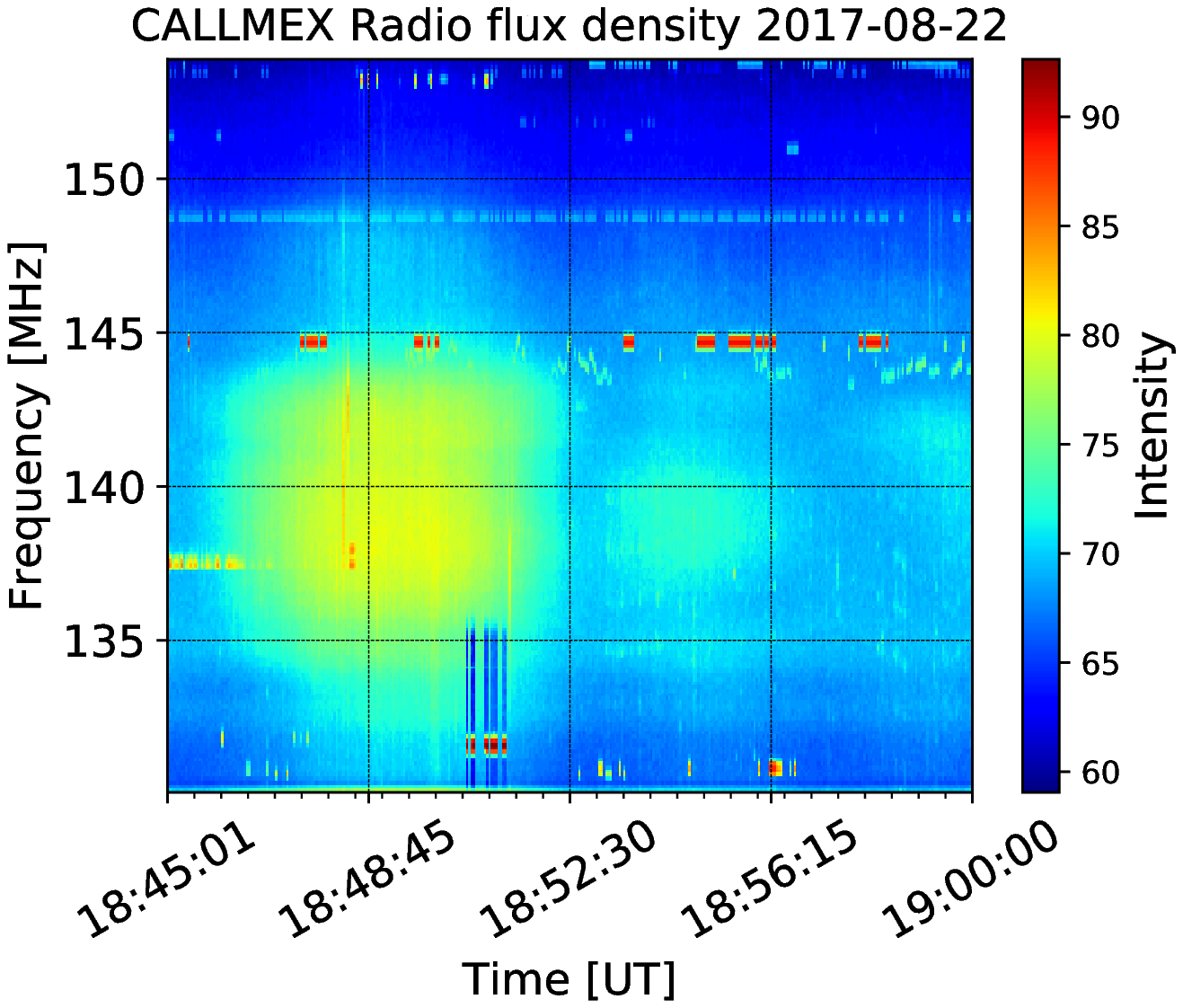}
  \end{tabular}
  \label{spectrums}
\end{figure}

Much spontaneous noises come from electronic devices, weather activity, power transmission networks, radio broadcast, etc. Looking at an instant spectrum of CALLMEX, we find these main sources of noise are present over the background as the signal with the lowest intensity. Stronger signals such as radio and TV broadcast range from 50\,MHz to 110\,MHz, as seen 
in the first spectrum of Figure \ref{spectrums}; finally, a lower noise component is present from 110\,MHz to higher frequencies. 

One commonly appearing source is the zebra-pattern noise; typically it occurs in dynamic spectra as parallel stripes oscillating in frequency over time. This kind of emission can be produced by natural means, with different mechanisms that have been proposed. The source of such emission is located in the solar atmosphere at the eruptive event's environments as described by \cite{altyntsev2005origin}, \cite{ledenev2006interference} and \cite{chernov2018model}. Nevertheless, another cause of this type of radio signatures is due to electronic-device inference near the instrument of detection, as signalized in the common radio signals catalog of \cite{monstein2011catalog}. The zebra pattern in CALLMEX is mainly caused by local electronic interference as shown in the second spectrum of Figure \ref{spectrums}.

Some emissions of natural origin detected by CALLISTO include lightning flashes; on the dynamic spectra they are very intense pulses with a duration of one second or less, but they can cover a large spectral range. The whole spectra can be crammed precluding the analysis of signals of interest during intense lightning activity. Electric storms are presented in this study as noise 
radio-interference patterns in both CALLMEX and MEXART signals as shown in the third spectrum of Figure \ref{spectrums}. 
One feature of CALLISTO is its flexibility of using many types of antennas for its receiver, in one experiment we used one line of antennas of MEXART and connected it to the CALLMEX receiver to detect the transit of the Sun as shown in the last spectrum of Figure \ref{spectrums}. In this spectrum, the big bright spot indicates the transit of the Sun over the main beam, which lasts around eight minutes, the next spot in 18:51:01 is the lateral beam of the radiation pattern of the antenna detecting solar emission.

\subsection{Solar Radio Burst (SRB)}

In February 2015 CALLMEX started to record data to detect SRB. 
The first scientific measurements occurred on 29 September 2015, when a SRB of Type III was detected, showing its characteristic slow drift in frequencies, along with a broadcast signal interference of short duration (as seen in Type III, in the third row of Figure \ref{mosaicII}).

During the next months, CALLMEX detected SRB Type I, II, III, and V, with Type III the most frequent.
The observed Type II SRB presented their typical signature: frequency drift and sometimes its first harmonic. These events are 
a noise source for MEXART IPS observations.
Once the events are detected by CALLMEX and then validated with the CALLISTO network, they can be now be identified in the MEXART data. 


\subsection{Radio Background}
During some solar events, deep sky sources transited through the MEXART beam. In order to explain the signal-background profile, we noticed that the MEXART beam width ($\approx 1^{\circ}$) covers $0^{\circ}\,-\,4^{\circ}$  
in the right ascension direction; thus radio source transit occurred at this timescale. During this work, we identified four events whose background signal was related with known celestial sources transiting through the MEXART beam and exceeding the limiting sensitivity value of 25\,Jy. 

From this, the Cassiopeia A supernova remnant transited during the 26 May 2016 event. The background variation was well observed with an amplitude $\approx 10^4$\,Jy \cite{1998BSAO...46...62T}, comparable with those of the event.

During the event of 8 September 2017 a supergiant elliptical galaxy (M87) and a BL Lacertae object (3C 273) contributed to background signal with $\approx 10^3$\,Jy \cite{2010A&A...511A..53V} and $\approx 40$\,Jy \cite{2010A&A...511A..53V} respectively. 
In this case the background amplitude was much less than that of the event.

Cygnus A is a Seyfert-2 galaxy, which transited the 2 February 2018 SRB event. This source has an estimated signal of $\approx 10^4$\,Jy \cite{2010A&A...511A..53V}. The basal flux registered variations of the order of four minutes from the collective signal of the pulsars G063.7+1.1, G065.7+1.2, G067.7+1.8, and G069.7+1.0, all of them with a flux less than 10\,Jy \cite{1998BSAO...46...62T}

Finally, on the 6 May 2019 20:45, an event occurred during the incursion of radio source G160.9+2.6 in to the MEXART beam. This source has an estimated flux of 350\,Jy. Later on the same day, another event occurred at 22:18:30, at this time, the sources G189.1+3.0 and G192.8-1.1 transited through the MEXART beam giving an estimated total flux of 430\,Jy  \cite{1998BSAO...46...62T}


In the following section we describe the joint SRB observations of MEXART and CALLMEX instruments.

\section{Joint Observations}\label{observations}
To assure the identification of SRB in our instruments, we first used CALLMEX to identify intense events. Where doubtful emissions are, the spectra are compared with the data from the e-CALLISTO network to find the same features at the same ranges of frequencies and times. If the SRB is confirmed by other stations in the e-CALLISTO network then we search for the event in the MEXART data.
This methodology helps to identify strong SRB in the MEXART data using CALLMEX records directly. However, for the case where the bandwidth in CALLMEX does not cover the event, it is necessary to use another e-CALLISTO station to corroborate the SRB.

In the following subsections we describe the characteristics of the signal recorded in our instruments per SRB type during the period of observations.
\begin{figure}
  \centering
  \caption{\textit{Left}: Different SRB types spectra detected by CALLMEX (I,II,III, and V in the final row). \textit{Right}:  In the right column, the same events, observed with other CALLISTO stations are displayed.}
  \begin{tabular}{cc}
    \includegraphics[width=56mm]{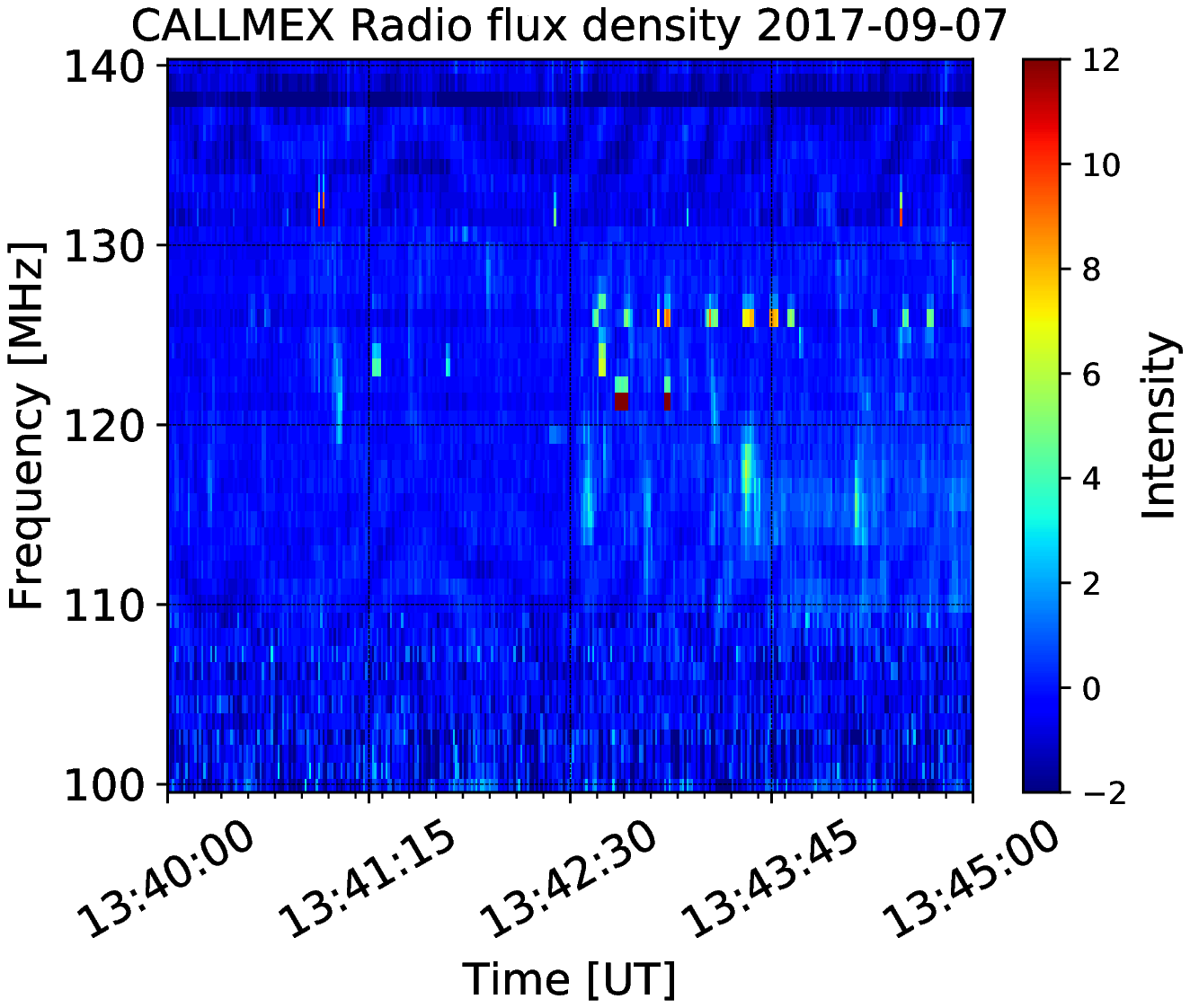}&
    \includegraphics[width=56mm]{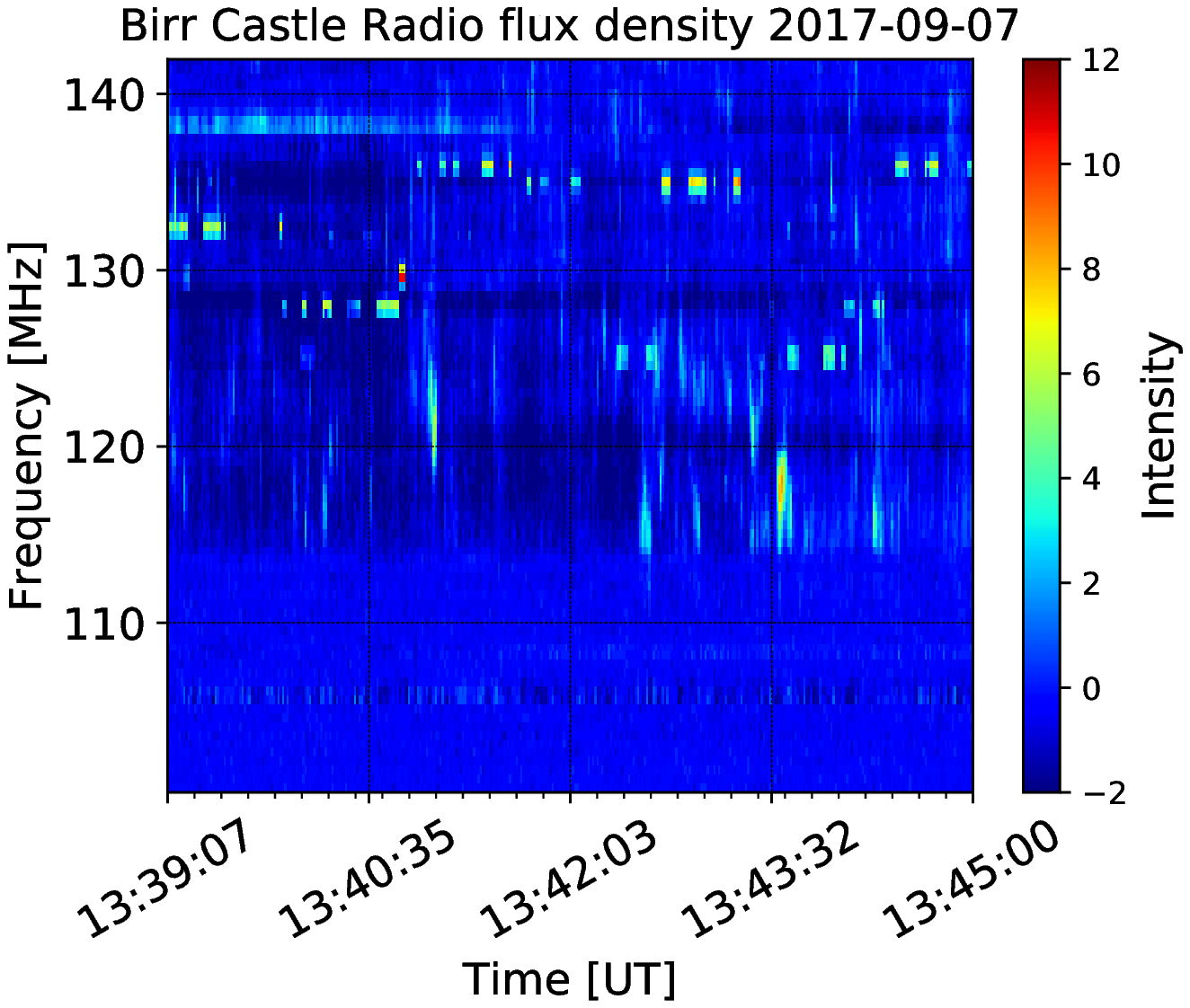}\\
    \includegraphics[width=56mm]{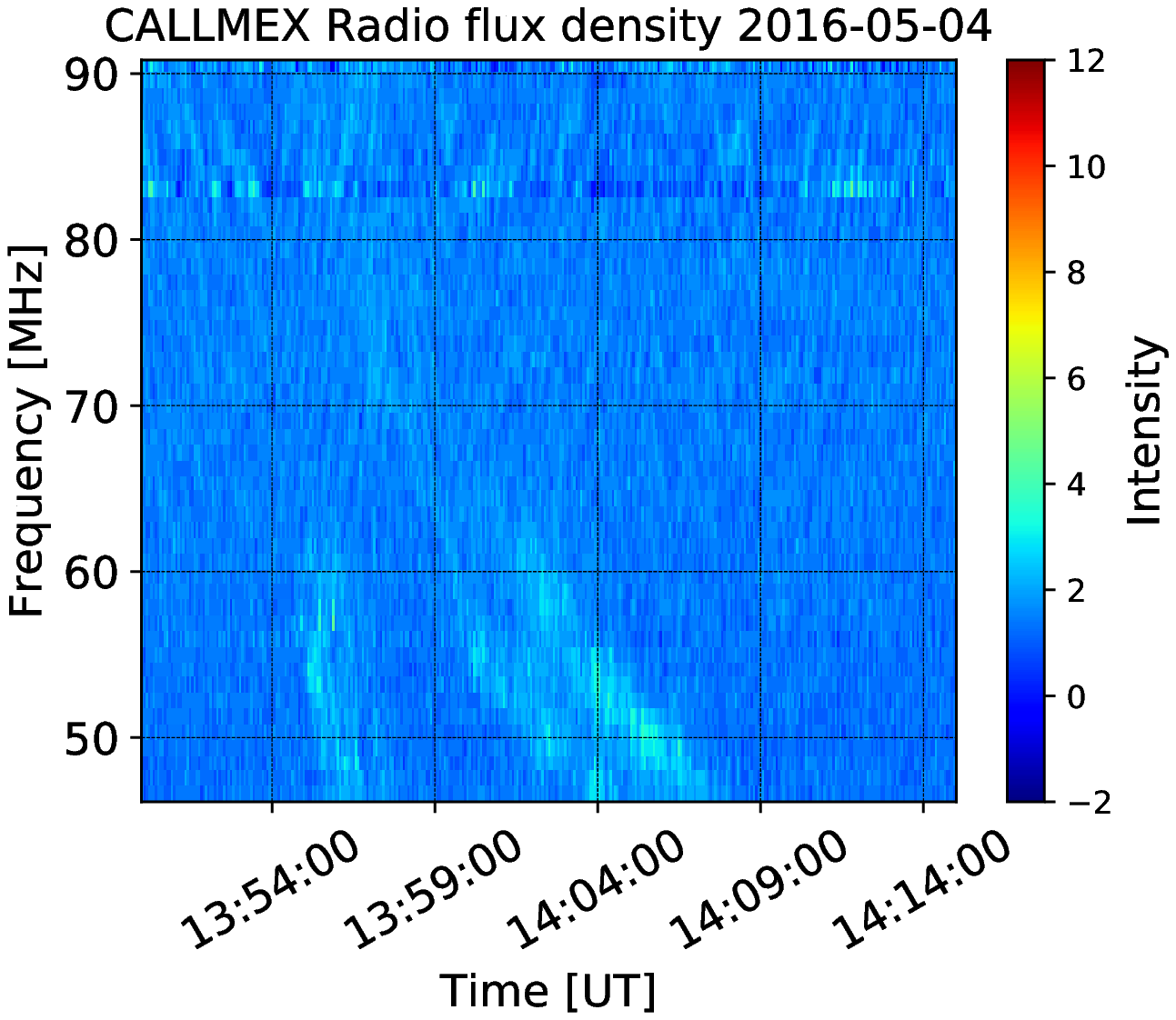}&
    \includegraphics[width=56mm]{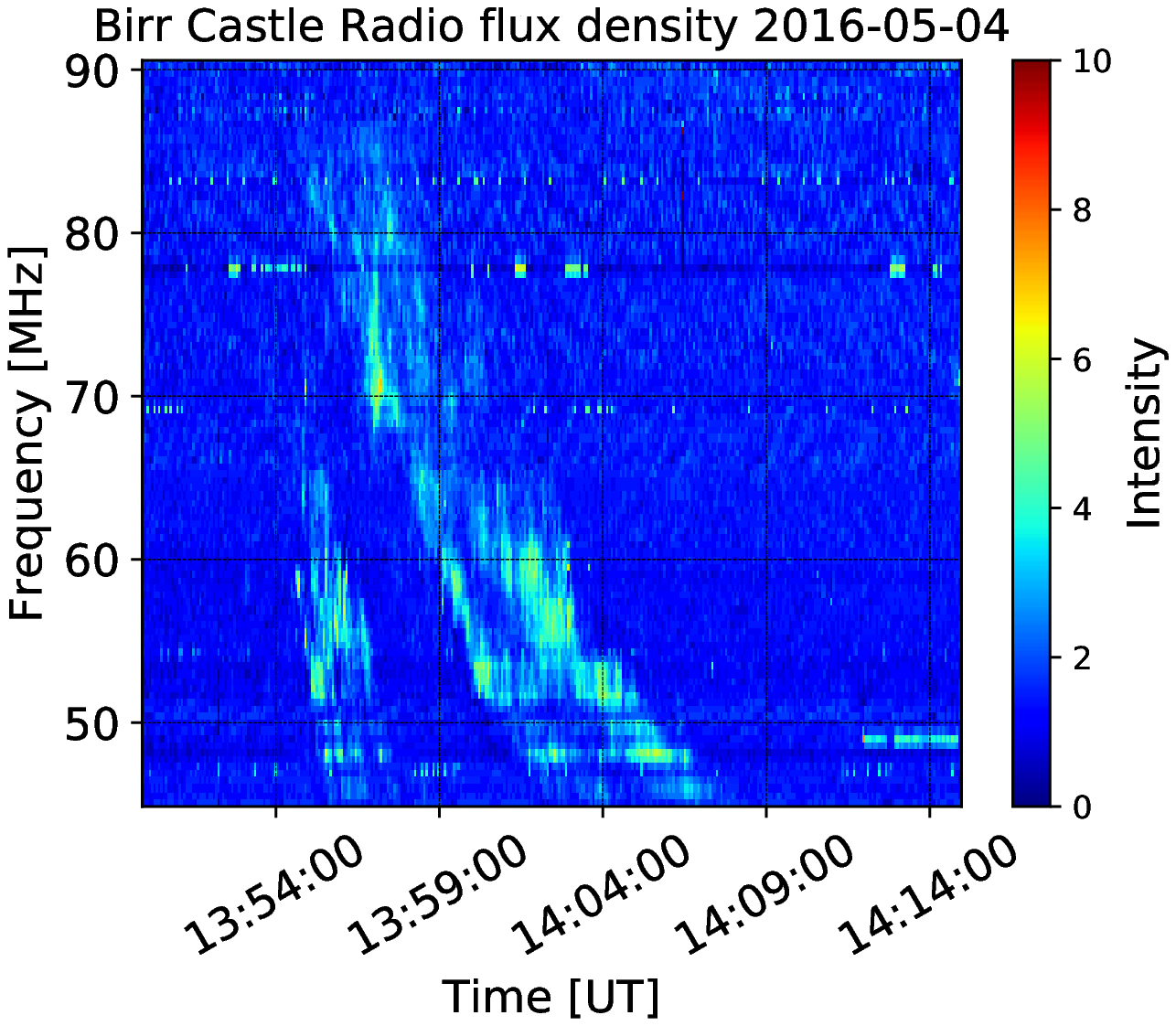}\\
    \includegraphics[width=56mm]{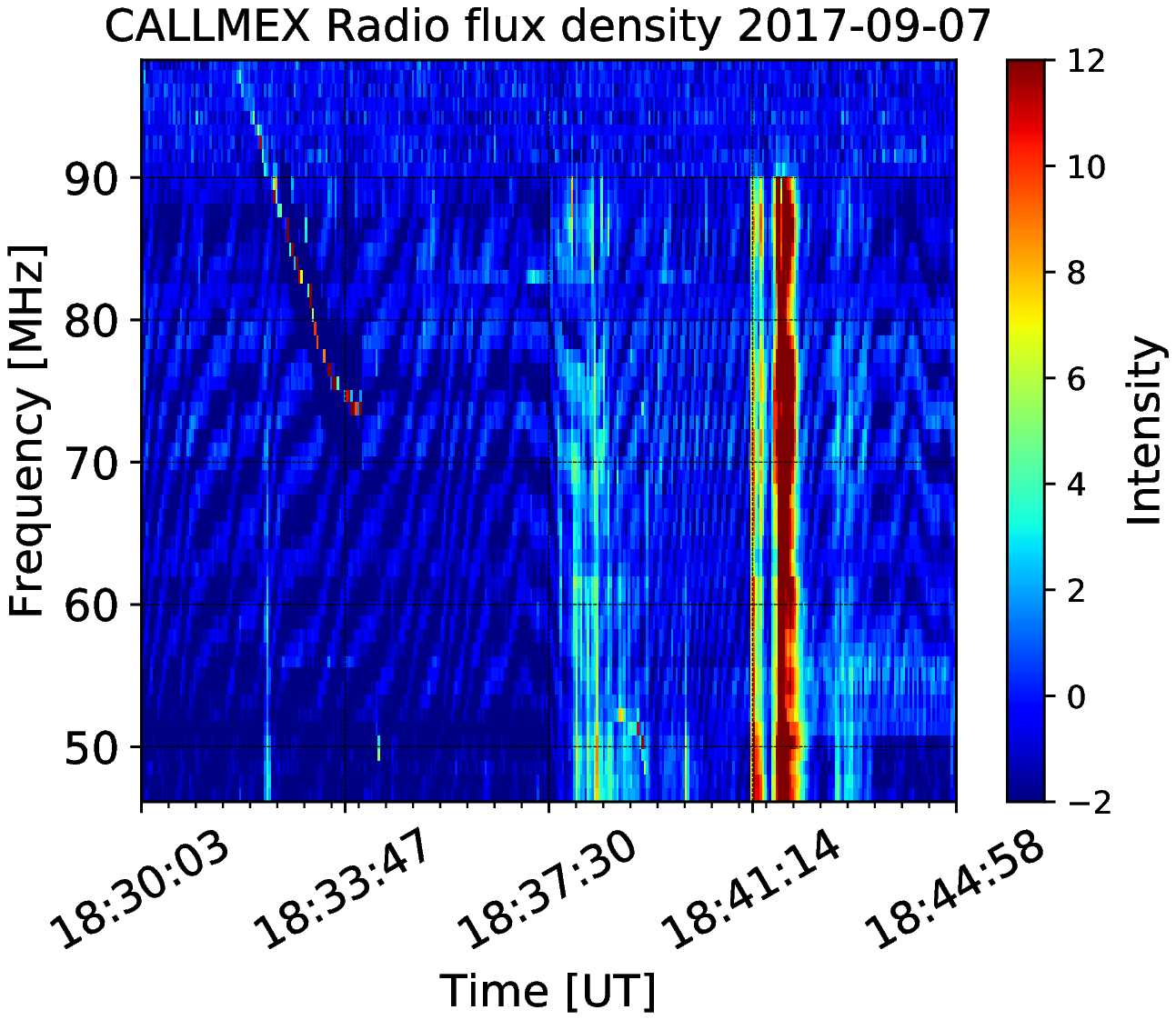}&
    \includegraphics[width=56mm]{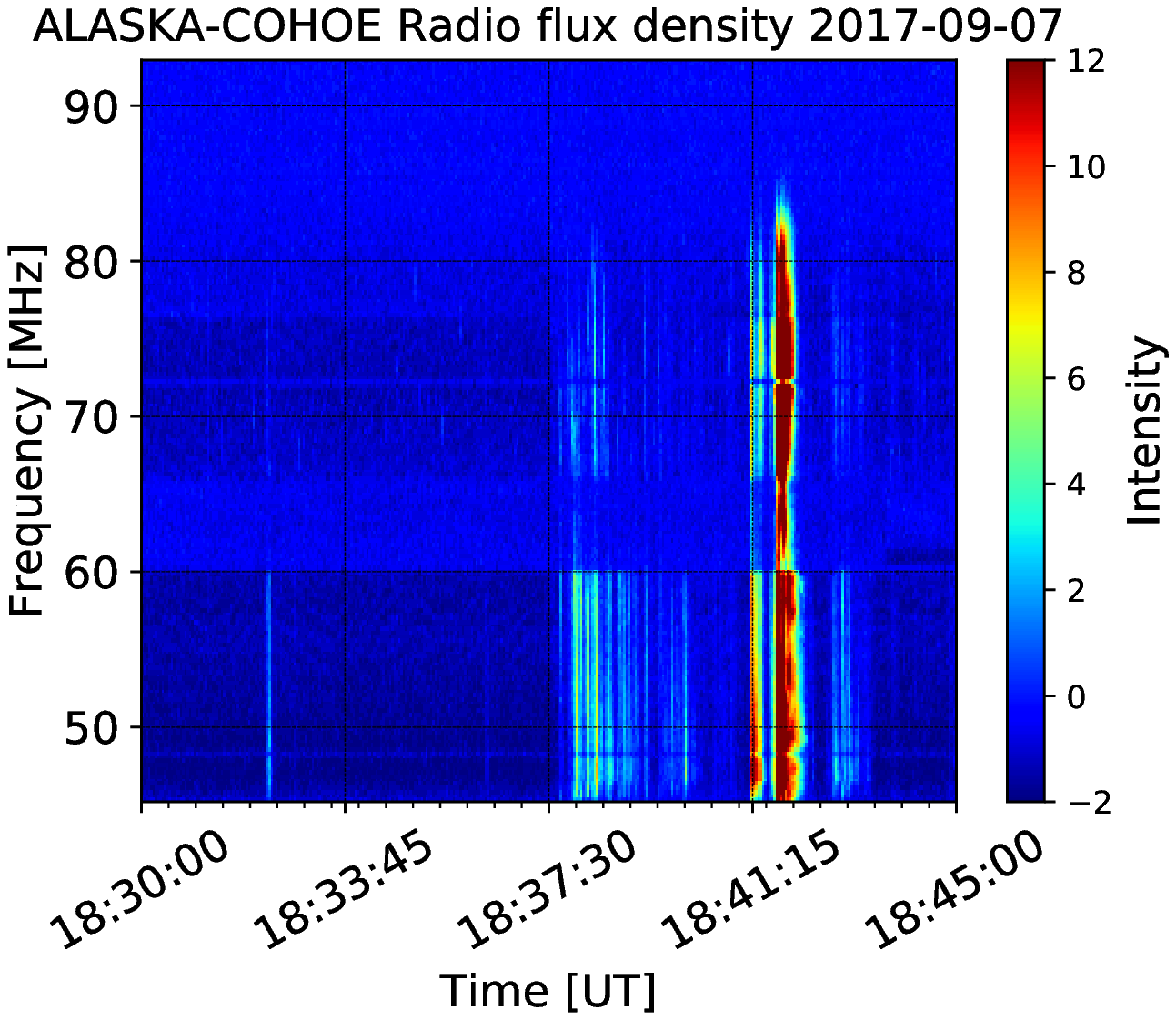}\\
    \includegraphics[width=56mm]{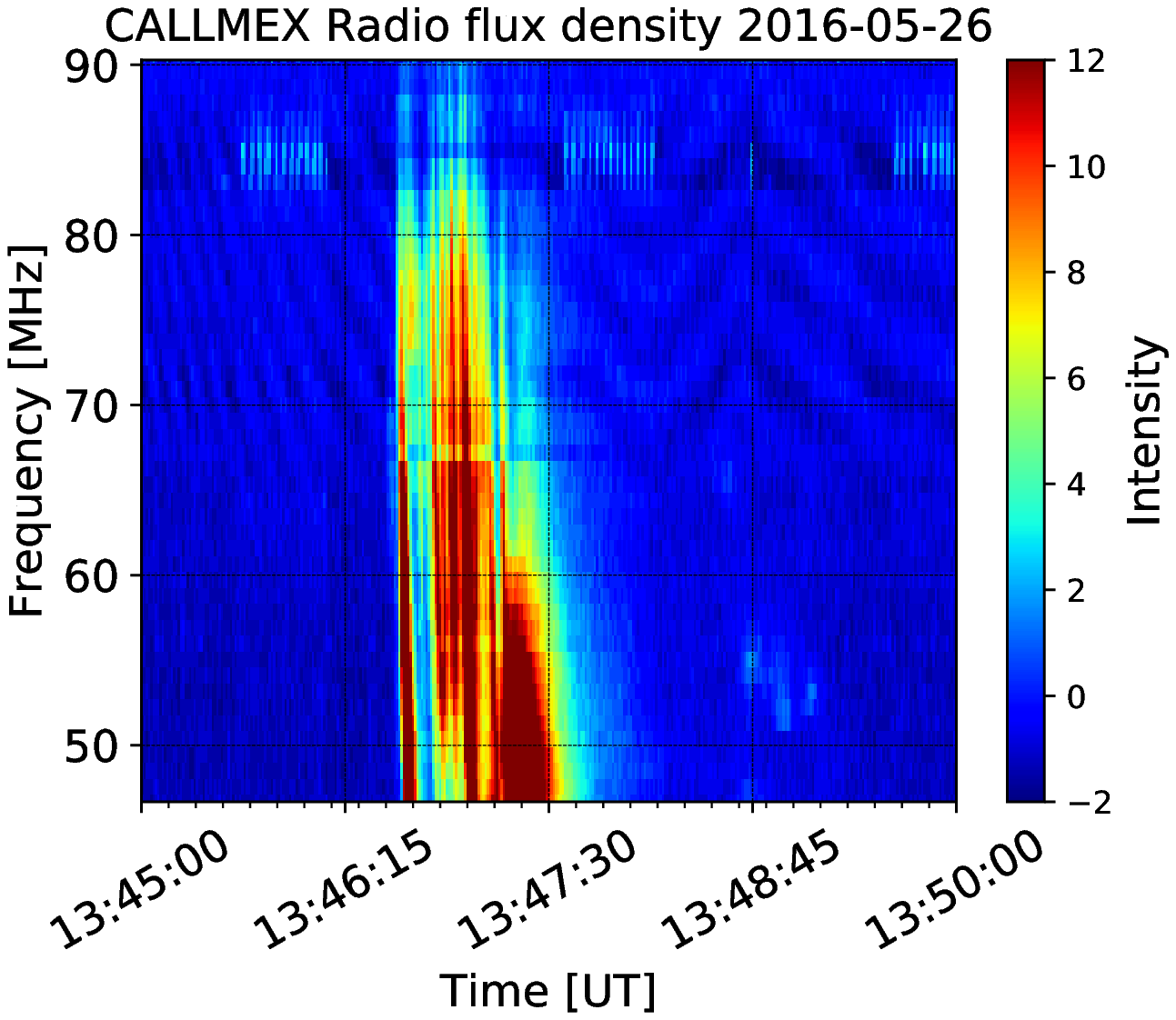}&
    \includegraphics[width=56mm]{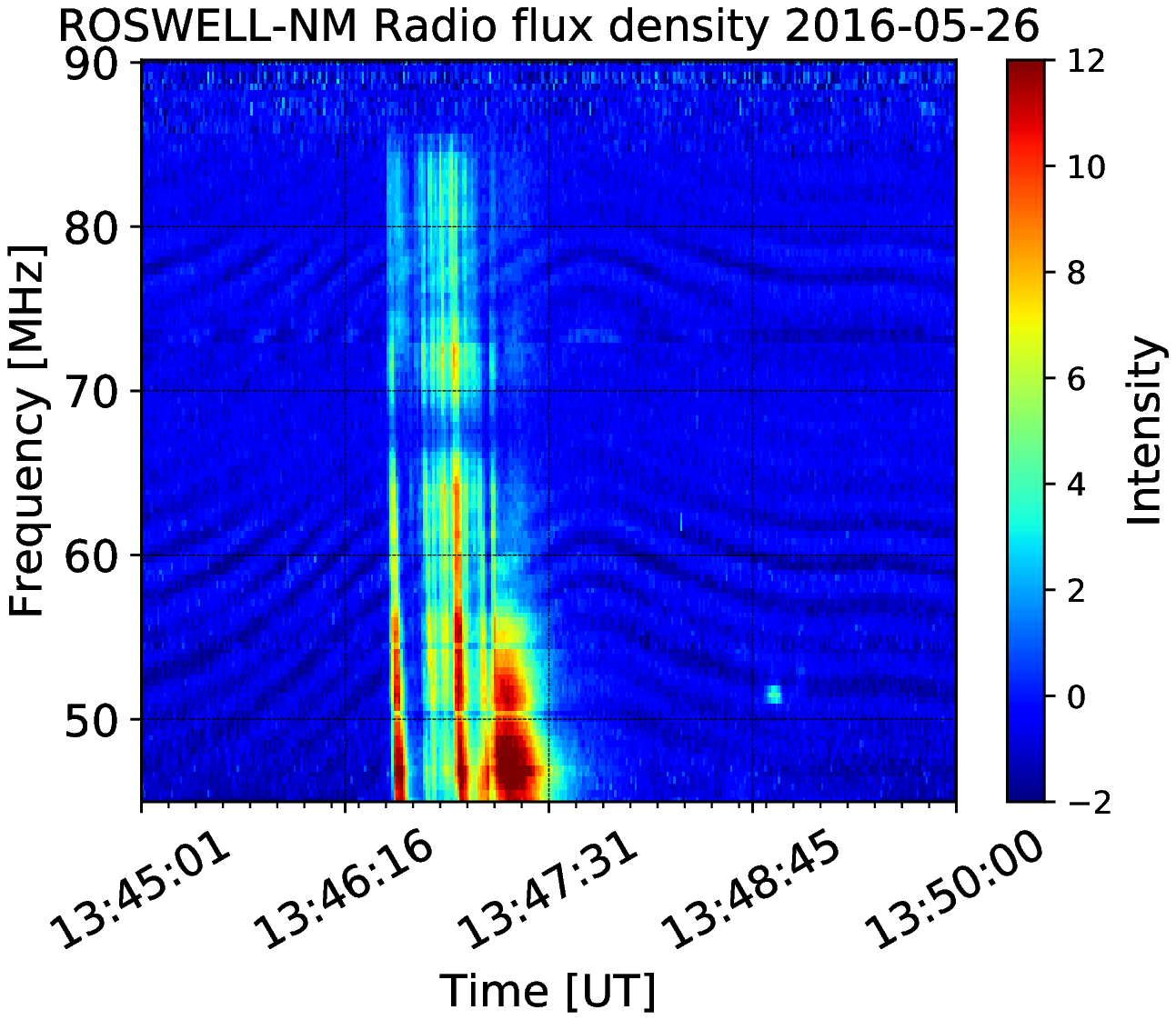}\\
  \end{tabular}
\label{mosaic}
\end{figure}
\subsection{SRB Type I} %
During the whole period of CALLMEX observations reported in this article, just one single Type I SRB was detected on 7 September 2017. This event presented the lowest intensity level observed for a SRB, just over the background noise level. Its spectra was compared with the Birr Castle station (see first row of Figure \ref{mosaic}) at spectral range
similar to that of the CALLMEX station.
As a verification, three intense pulses were compared, one at 13:41:04 for 120\,MHz, the second at 13:42:36 for 115\,MHz and the third at 13:43:36 at 117\,MHz. These three and other signals were jointly registered by the two instruments separated by more than $6000$ kilometers. 
MEXART was operating during the event but in this case, the signal of the SRB is not clear enough to be characterized with both instruments at 140\,MHz as seen in the first lightcurve of Figure \ref{mosaicII}, MEXART detects peaks of noise, not related with the Type I signal detected with CALLMEX, and  a increasing flux, possibly from a distant radio source. 
 The range of time  compared with MEXART goes from 13:40 to 13:45 UT; nonetheless, this event lasted for almost ten hours, and other Type III events occurred during this time.

\subsection{SRB Type II}

We identify seven SRB of Type II during the period of observation.
One of the first Type II events was detected on May 2016 (second row of Figure \ref{mosaic}); CALLMEX detected the event in a noisy zebra background. The emission started at 13:54:39 and ended  at 14:07:44; the initial frequency of emission was 86\,MHz, as this type of events drifted to lower frequencies, CALLMEX missed the final frequency of emission as this instrument only reach 45\,MHz. As we can see, there are two main bands of emission; a higher band seem to be an harmonic component of the lower band as it  doubles the frequency of emission. To verify this emission the event is compared in this case with the BirCastle station, this station has a narrower bandwidth so it detects the event with more detail than CALLMEX, but as we can see, the emission is identical in frequencies and time.

A second case, the event of 10 July 2016, (see the second row of Figure \ref{mosaicII});  started 
at around 01:00 UT at a frequency higher than the bandwidth of CALLMEX; the emission continued below the lowest detection frequency of 45\,MHz. As the burst drifted to lower frequencies, MEXART started to detect the emission 24 seconds later in a noisy background and stopped detecting at 01:02:23; the emission was detected for 7 minutes and 9 seconds in CALLMEX with its full bandwidth and 1 minute and 59 seconds with MEXART, as seen in the second lightcurve.

\subsection{SRB Type III}
The event of 7 September 2017 (third row of Figure \ref{mosaic}) started at 18:32 with a faint emission covering a band observed from 90\,MHz to 45\,MHz. Afterwards, the emission exhibits a chain of pulses at 18:37, growing fainter, but with a final bright pulse at 18:41. This final pulse also thickened resulting in a Type V emission, and ended with fainter pulses at 18:43. This event is compared with the Cohoe Alaska station, which has lower background noise and also more definition due to its narrower band (90\,--\,45\,MHz).

\begin{figure}
  \centering
  \caption{\textit{Left:} The spectra of different types of SRB, detected by CALLMEX, are displayed. \textit{Right}: The same events are plotted with the MEXART timeseries and the 140\,MHz CALLMEX channel. The intensity is presented as voltage detected by the antennas receivers with different sensitivity and both signals amplitude were scaled arbitrarily to match the signals.}
  \begin{tabular}{cc}
    \includegraphics[width=56mm]{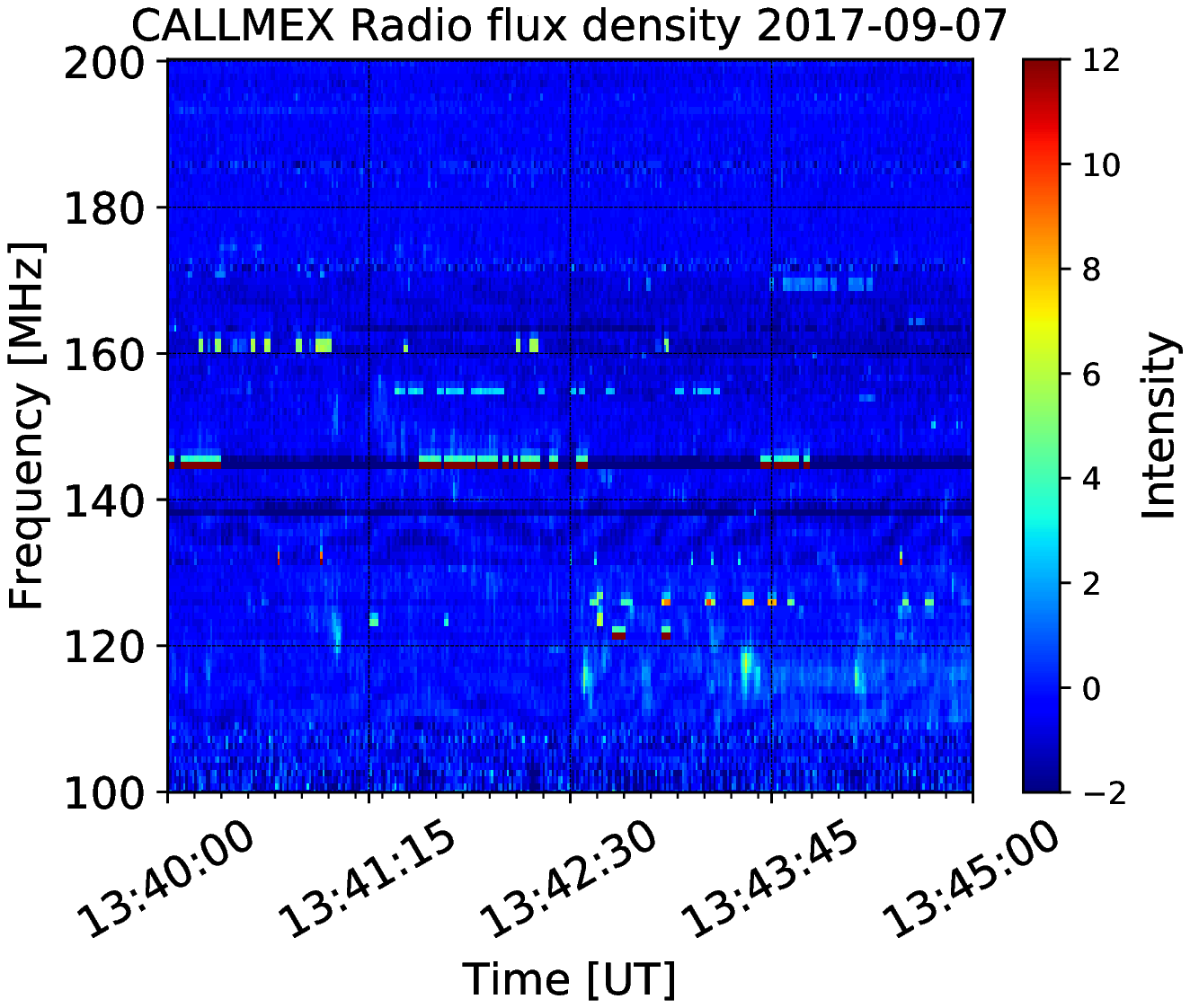}&
    \includegraphics[width=56mm]{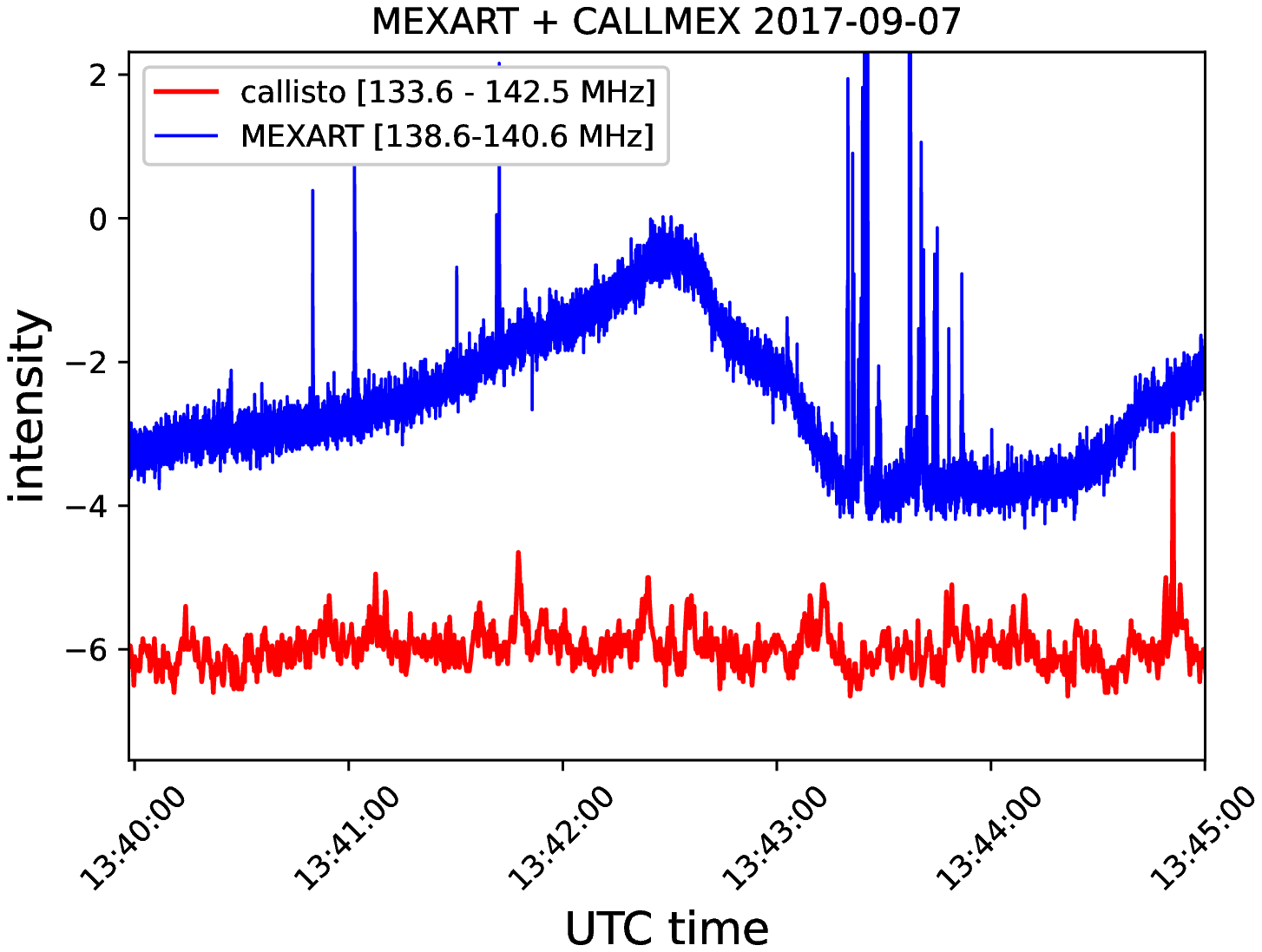}\\
    \includegraphics[width=56mm]{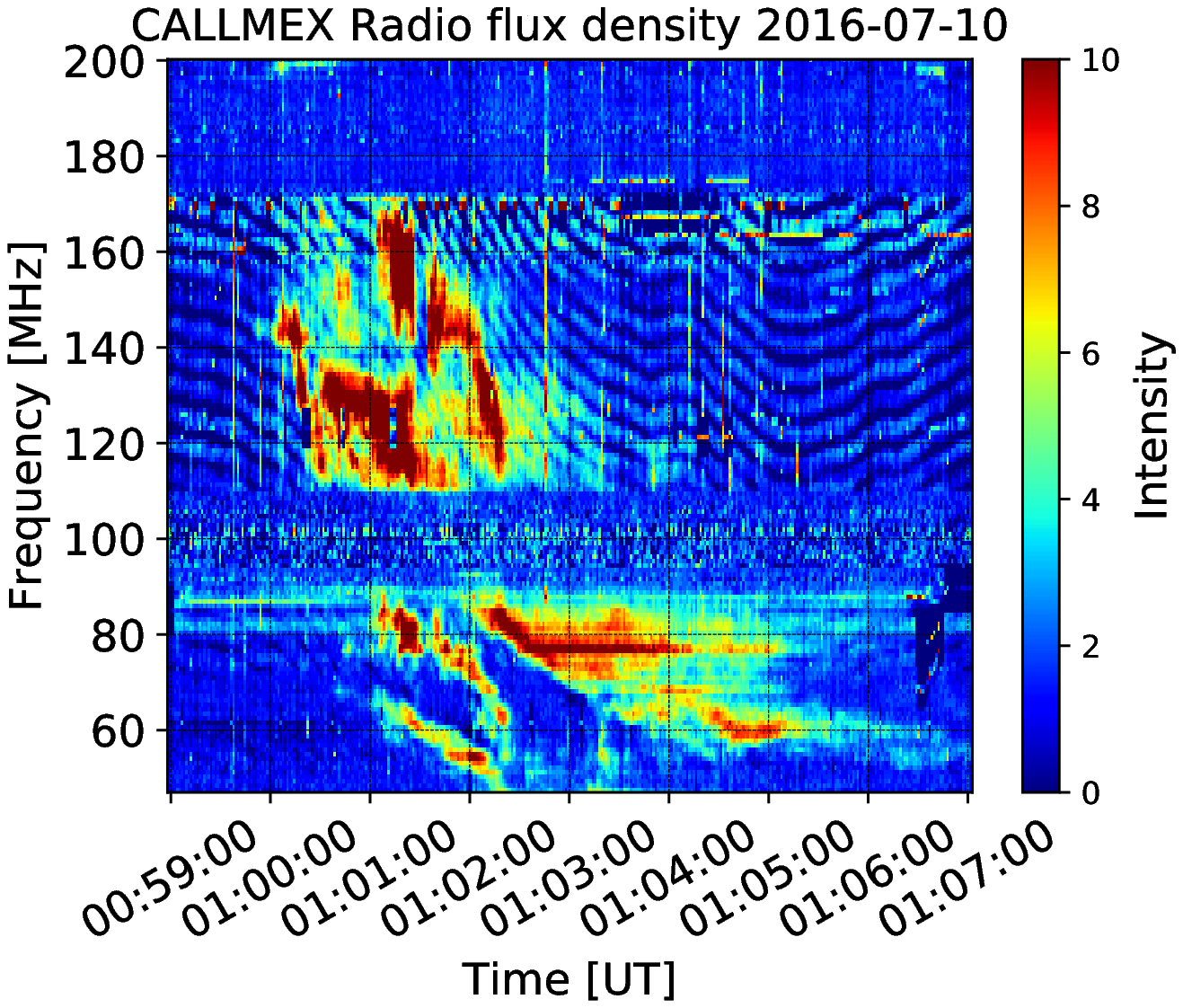}&
    \includegraphics[width=56mm]{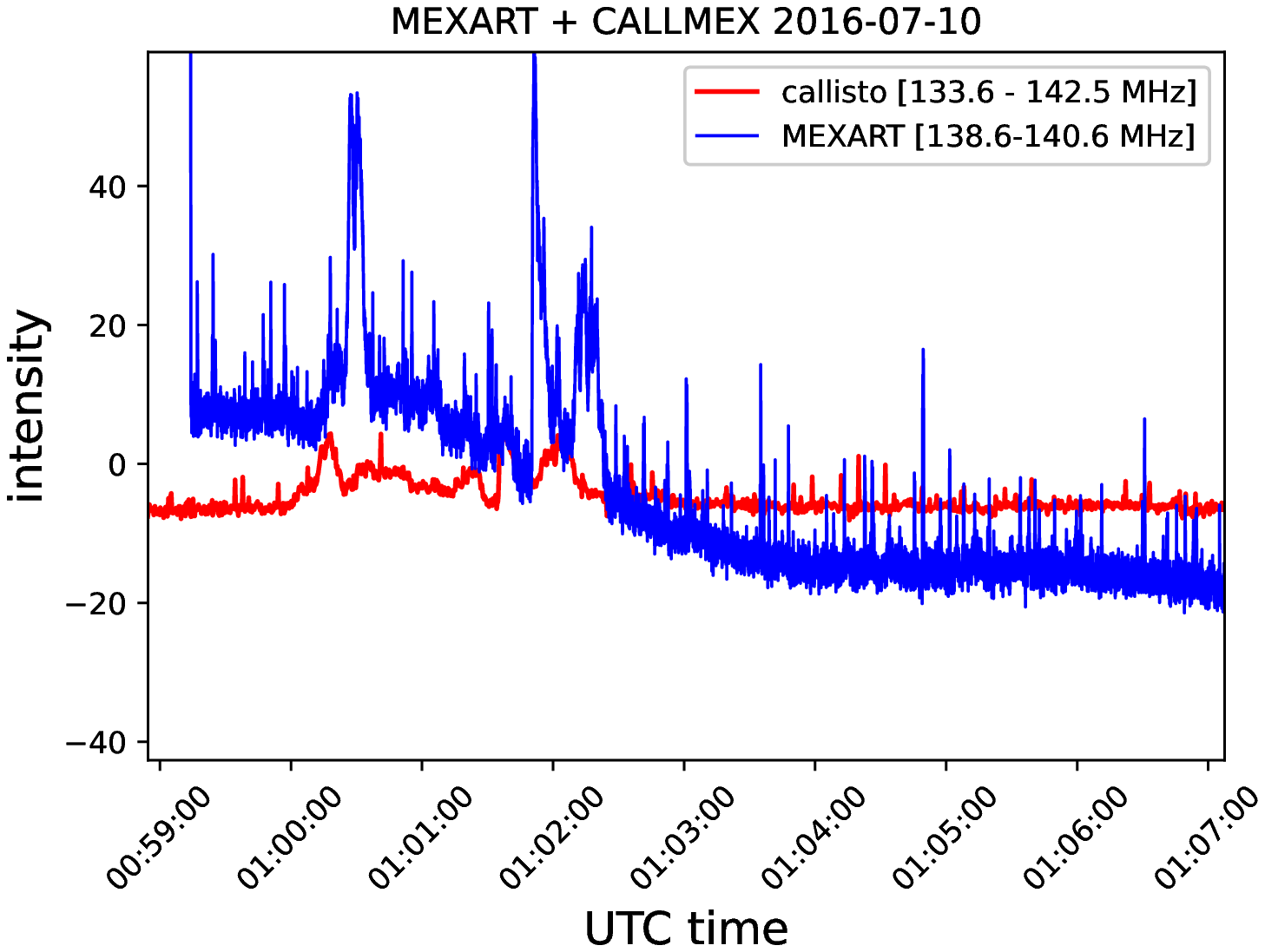}\\
    \includegraphics[width=56mm]{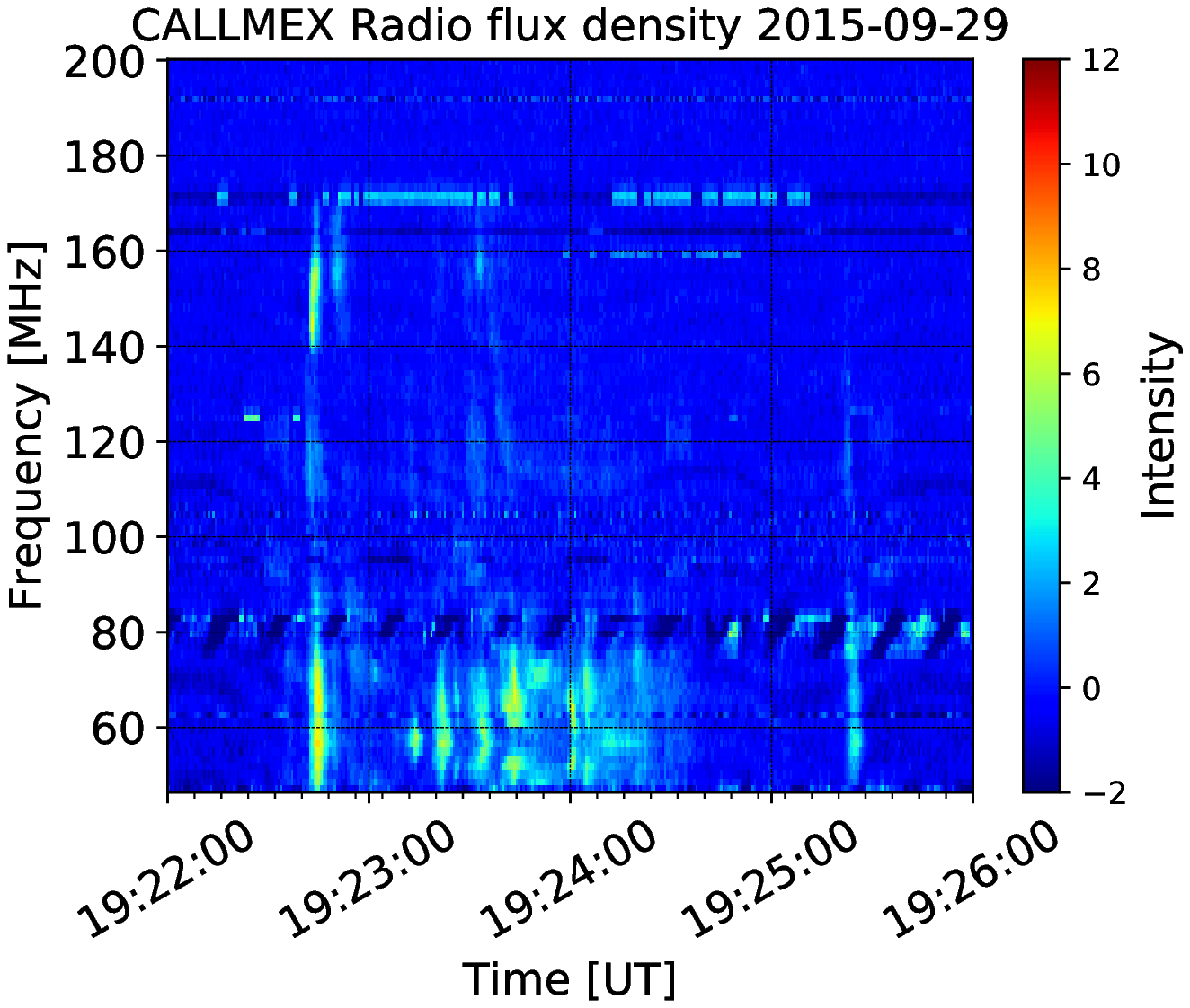}&
    \includegraphics[width=56mm]{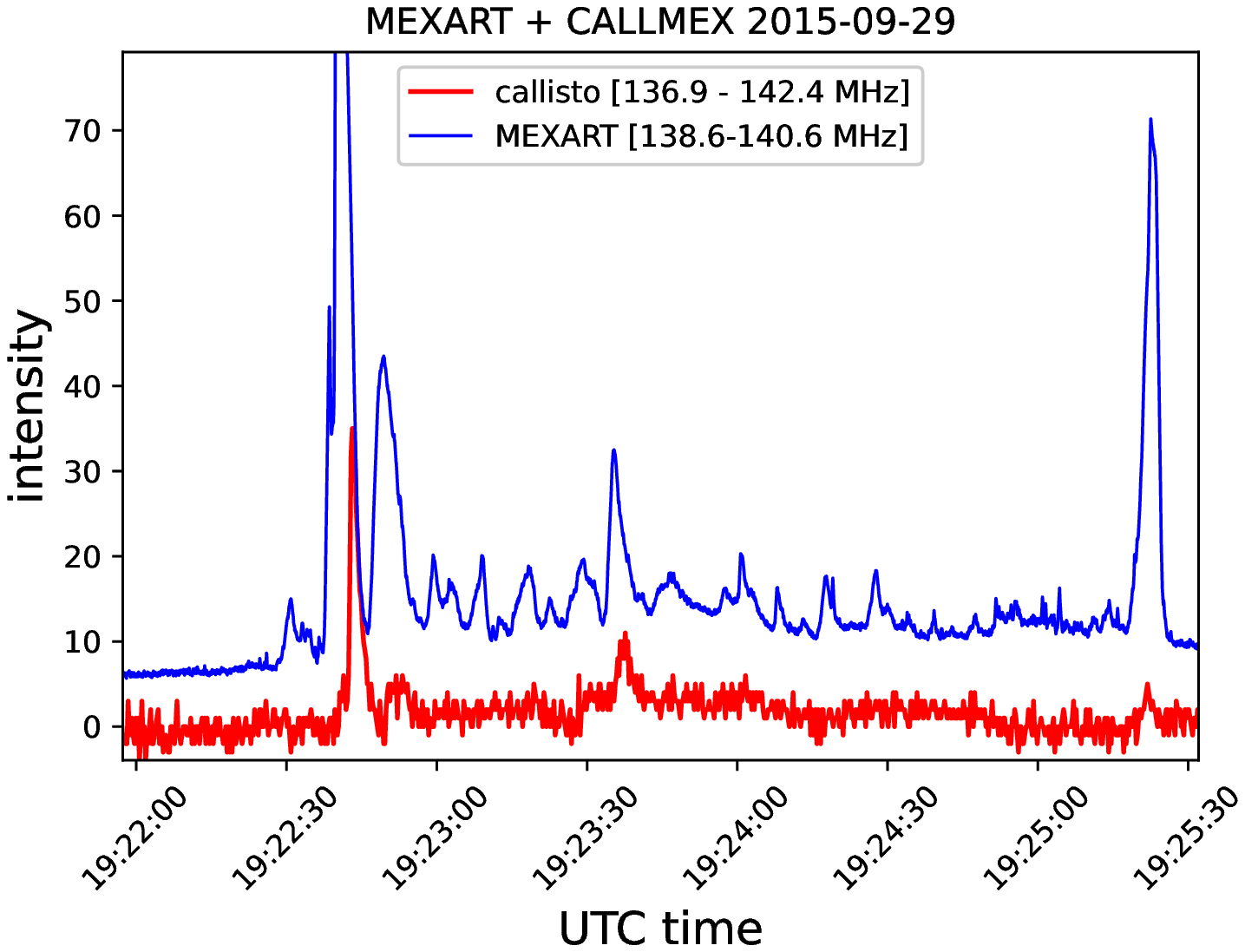}\\
    \includegraphics[width=56mm]{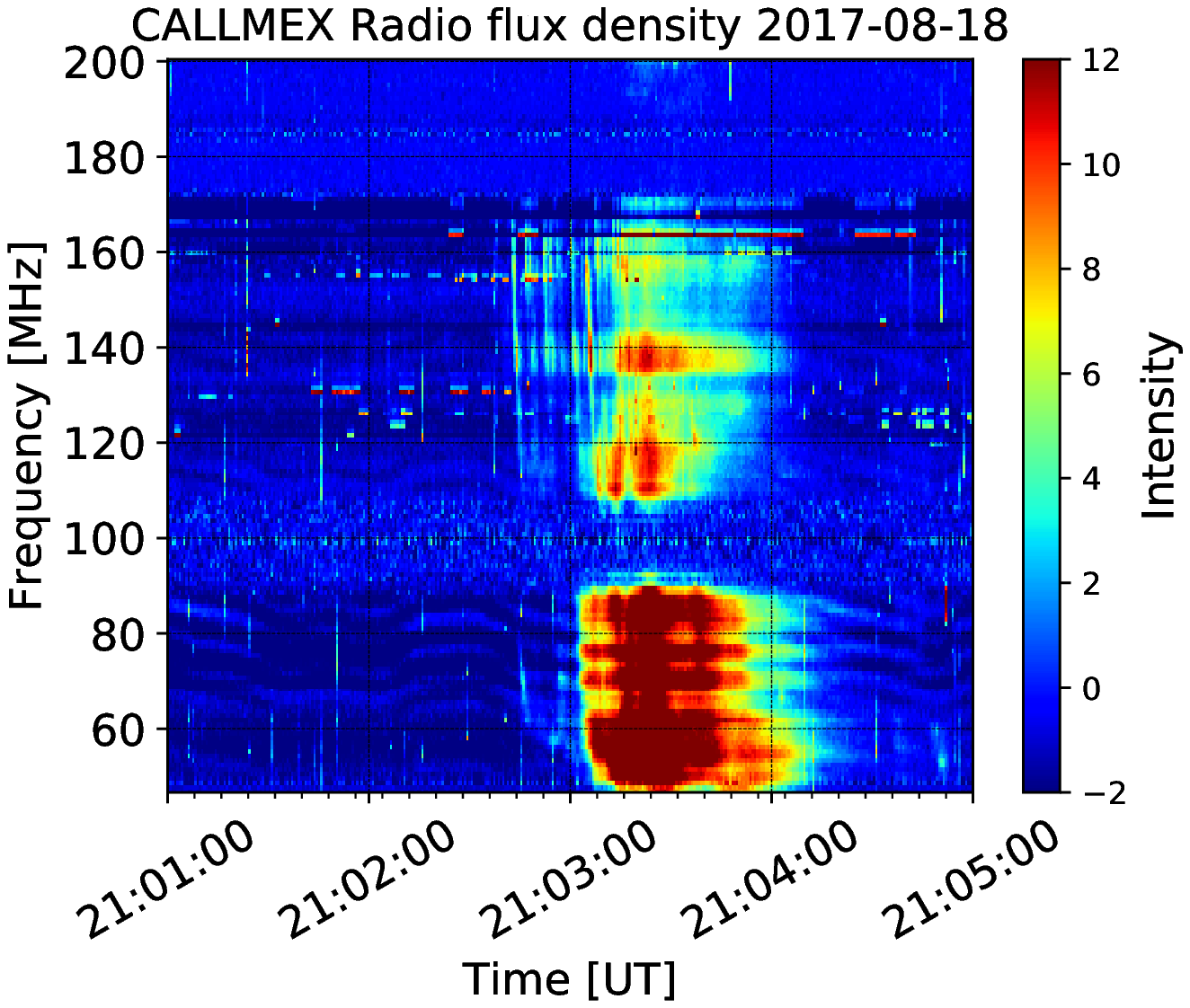}&
    \includegraphics[width=56mm]{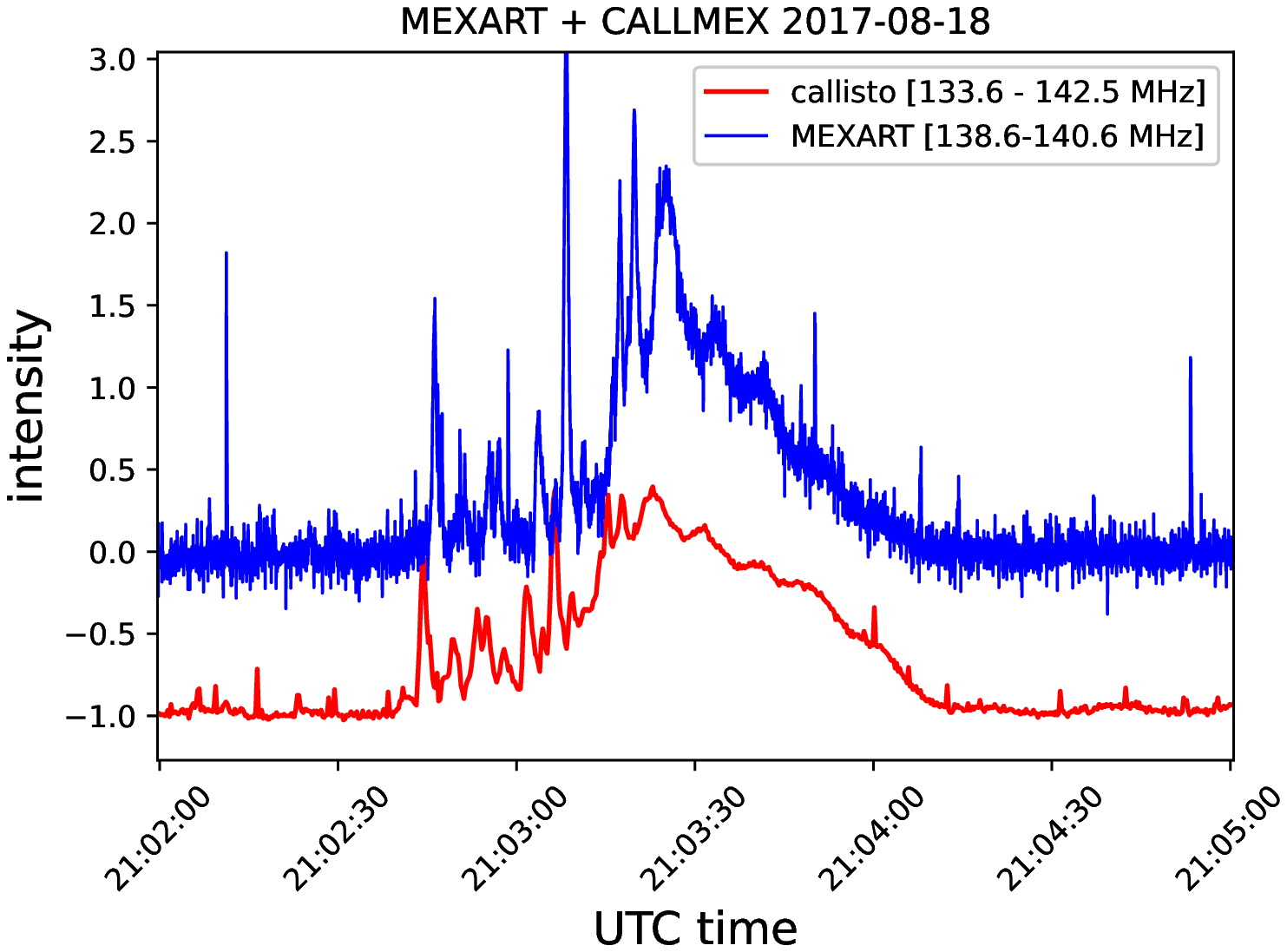}\\
  \end{tabular}
\label{mosaicII}
\end{figure}

The event of 29 September 2015 was the first light of CALLMEX, a SRB Type III (third row of Figure \ref{mosaicII}); its detection in CALLMEX started at 19:22:34 and lasted 
2 minutes and 53 seconds. However, MEXART started its detection 7 seconds earlier, in fact, MEXART  detected the initial peak with a 629 signal to noise ratio (SNR) while CALLMEX detected the same peak with a 14 SNR in the same frequency channels where MEXART operated,  this demonstrates the higher sensitivity of MEXART. Also from the spectrum it can be seen that the SRB had many stages of emission, first at 19:22:44 occurred a bright emission with a wide bandwidth, from 45 to 170\,MHz. In a second stage, from 19:23:12 to 19:24:36, a fainter emission was present mostly in a bandwidth from 45 to 80\,MHz. Lastly, the emission seems to fade completely, but it ends with a strong brightening at 19:25:23 with a similar bandwidth to the first stage. This behaviour is present also in the 140\,MHz band, where the initial and final pulse is present and also the intermediate and fainting stage. This stages feature was noted in many Type III events, so this is analyzed more rigorously in a later section.

\subsection{SRB Type V}

Some other feature detected in the Type III events were the banner type emission, identified as the Type V emission.

One of these events detected with CALLMEX is that of 26 May 2016 (last row of Figure \ref{mosaic}), CALLMEX detected the emission at 13:46:34 and the emission immediately thickened producing a Type V emission. The event compounded many thick pulses and ends at 13:47:57.

To verify this event, the Roswell station is used; we can see that sensitivity in the Roswell case seems to be lower than in CALLMEX.

In the final row of Figure \ref{mosaicII} we present a Type V event as the product of a Type III event. In this case, the event started with a Type III emission, at 21:02:43 with Type III spikes, at 21:03:02 the emission changed in its shape as it thickens toward lower frequencies and has a brighter emission. This banner acquired shape is recognized as the Type V emission. MEXART also detected the event; in this case,  the emission started with narrow spikes, and after some seconds the emission became more continuous and brighter; in the late stage of the event, the emission faded out.

\subsection{Directivity Analysis}




While reviewing the different SRB detections with MEXART, it was noted that the position of the Sun during these events were not only in the local meridian as expected from MEXART's directivity for signal detection, but they are cases in many altitudes, including near to the horizon.

\begin{figure}  
    \centerline{
    \includegraphics[width=1\textwidth]{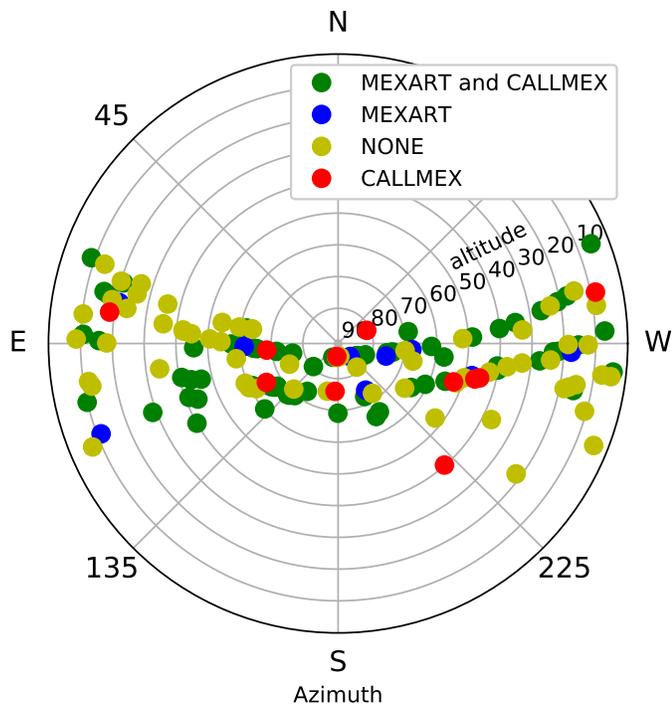}
    }
    \caption{Polar projection of the sky in horizontal coordinates locating the location of the Sun during the occurrence of the burst detected by CALLMEX. The green dots represent the cases where MEXART and CALLMEX detected the event at 140\,MHz, the blue dots represent the cases detected at 140 MHz only by MEXART, the yellow events indicate the events where neither MEXART nor CALLMEX detected the burst at 140\,MHz and the red dots represent the cases where CALLMEX detects the event at 140 MHz but there is no detection by MEXART. }
    \label{polar}
\end{figure}

In Figure \ref{polar}, we plot the Sun position in the local horizontal  coordinates space 
for all recorded events in the projected sky. 
The green dots represent the events detected by MEXART and CALLMEX, the blue dots are the events detected only by MEXART at 140\,MHz, the red points represents the events observed in 140\,MHz only by CALLMEX, and the yellow dots represent the events where neither MEXART nor CALLMEX detected the event in the 140\,MHz channel. In total, there are 151 events where MEXART was operational. These yellow cases may represent one of many situations as reason for the lack of detection of the event at 140\,MHz. One of them is that the SRB simply had no detectable 140\,MHz emission, another reason could be that the intensity of the emission was to low to be detected by both instruments. This could be due to directivity reasons, or in some cases the MEXART signal is too noisy to assert that a detection was made with MEXART.  We observe that the distribution of SRB cover the Sun's transit path and is independent on 
the MEXART main beam pattern direction, which is focused along the local meridian. 
%


This SRB distribution shows that MEXART detects SRB isotropically, independently of the time of solar transit. With this distribution we show again that MEXART is sensitive in detecting SRB, as it detects some events that are not detected with the CALLMEX 140\,MHz channel.

\subsection{SRB Monitoring}
In order to ease the analysis of SRB in close to real time, we created %
three Python scripts 
using the Sunpy and Astropy packages \cite{stuart_mumford_2018_1160820,astropy:2018}. 
One program generates spectra with the CALLISTO data, the second program produces the CALLISTO lightcurves integrating the signal from different channels. Another script manages the flow of processes and distributes the new files and data files in a repository of files ordered by date.
Later; the lightcurves, spectra figures and the raw data  are inserted in the local data repository (\url{www.rice.unam.mx/callisto/}) and stored in the Mexican National Repository (\url{www.repositorionacionalcti.mx/}).
The raw data of CALLMEX are also stored in the e-CALLISTO database \url{www.e-callisto.org/Data/data.html} under the station code MEXART.

\section{Database of Solar Radio Bursts}\label{database}
The CALLMEX database 
includes each SRB event detected by CALLMEX and 
MEXART. 
The 
parameters obtained of the analysis from each event are summarised as follows:
\begin{itemize}
    \item Name of the event: in the format of YYYYMMDDhhmm indicating date and time of occurrence.
    \item Year, month, and date event for data querying.
    \item Time: of start and end of the event in UT.
    \item Type of burst.
    \item Bandwidth of frequency detected.
    \item Comments about the event. 
    \item URL: of the data file in the RICE repository.
    \item MEXART detection: with the values of NO or YES.
    \item RMS (Root Mean Square) and SNR calculated from CALLMEX lightcurve in the 140\,MHz band.
    \item RMS and SNR calculated from MEXART signal.
    \item Sun's position, at the moment of the event in horizontal and equatorial coordinates.
    \item Direction of the MEXART antenna pattern during the event.
\end{itemize}

A reduced version of the database can be found in table \ref{tableofallevents}.  The complete database is accessible at \url{www.rice.unam.mx/}

\subsection{Events}
The Table \ref{tableofallevents} shows the events detected by CALLMEX, in the 140 \,MHz channel, or by MEXART.  The SRB ID of the events tells the date and time of the event. There are two pairs of SRB IDs repeated in the table, this is because two different SRB types were reported at the same time, the event 201709121914 which presented a Type II and a Type III; and the event 210905071327 event with the same case. At the last part of the table it can be seen all the events Type III combined with the Type V characteristic. 
In total 156 radio bursts were recorded with CALLMEX, from which seven are Type II, one is Type I and 148 are Type III; from which 30 are accompanied by Type V emission. 
The reason for the big difference in the number of detections between the Type III events and the other type of events is mainly that the Type III are the most common of SRB \cite{Warmuth2005}. A second factor is the radio noise of the location, which has increased along the years. This affects mostly the detection of Type II events, which are fainter; and specially the Type I, which are the faintest of the SRBs and have a signature more difficult to identify over the local noise.  
From the 156 events, 75 were detected by MEXART; of these 75, CALLMEX detected the signal at 140\,MHz in 67 cases. In 13 cases only CALLMEX detected the emission at 140\,MHz when MEXART did not, in 5 cases MEXART was not operating,  and in the remaining 63  events, neither MEXART nor CALLMEX detected the event at 140\,MHz.

\subsection{SRB Related to Solar Flares and CMEs}

During the detection of SRBs, an event caught our attention; the Type II SRB with ID 201710202330 (see Table 2), two minutes earlier, at 23:28 occurred a M1.1 flare. Around 40 minutes later a CME was detected with the SOHO LASCO coronagraph with tag time 20171021T00:12. This event is an example that shows the relation between flares, SRB, and CME like was observed in other works \cite{lara2003statistical}.
A relationship was made between our SRB records and observed solar eruptive events. We used ”The Space Weather Database Of Notifications, Knowledge, Information” (DONKI) (\url{kauai.ccmc.gsfc.nasa.gov/DONKI/}), made by NASA's Community Coordinated Modeling Center. From this database, we looked at a range of time covering our SRB detections; from 2015 to 2019, where 624 CME were found at DONKI data. To relate Type II SRB we use the same criteria used in \cite{lara2003statistical}: search CMEs in a three hour window since the SRB was detected. We matched three Type II SRB with four different CME events (see Table \ref{typeiicme}). One SRB, the 201710202330 matched two different CMEs with the criteria mentioned before, also the SRB 201511041343 and 201710202330 have flares linked accordingly to the DONKI database. Later, for the Type III SRB, we searched all the flares in DONKI in the 2015\,--\,2019 years range and 150 flares events were available. In this case we fixed a window, three hours forward and three hours backwards to the SRB time, then we selected the flares which fitted the six hour window, see Table \ref{srbiiicmetable}. With this criteria, 33 SRB events were related with 23 different solar flares. Some SRB matched multiple flares; for example the SRB 201709042203f was related to four different flares, and in one of these flare cases with a CME and four Solar Energetic Particle (SEP) events as seen in the last page of the table. We can see that a great part of SRB are related with a CME and a flare simultaneously and all the flares have an intensity equal or greater than C1.5, just six SRBs Type III have no CME or SEP match, and only one Type II have no CME relation.



\section{Data Analysis}\label{dataanalysis}
\subsection{Phase Analysis}
The analysis of the data was accomplished by using radio-burst observations that were detected by the MEXART radio telescope previously identified in the CALLMEX station as Type III SRB. The analyzed radio bursts comprise 40 well defined events occurred at different epochs over the years 2015 to 2019. 

Observing the light curve of the analyzed radio bursts, it can be noticed that the whole structure of an event is is typically characterized by a fast increase of the intensity up to a maximum value, followed by a slow decrease in intensity. Also, comparing the results from the dynamic spectra and the corresponding light curves obtained from the observations, and based on the number of individual radio bursts that are present during an event, it could be possible to classify the events as single or composite ones. In this work single-radio-burst events are considered to be those constituted by one radio burst with a typical structure consisted of a previous, main, intermediate, and posterior stages. Composite radio burst events are considered to be constituted by multiple individual radio bursts occurring during a short interval of time, evolving during over few minutes.

\begin{figure}
    \centering
    \includegraphics[width=1\textwidth]{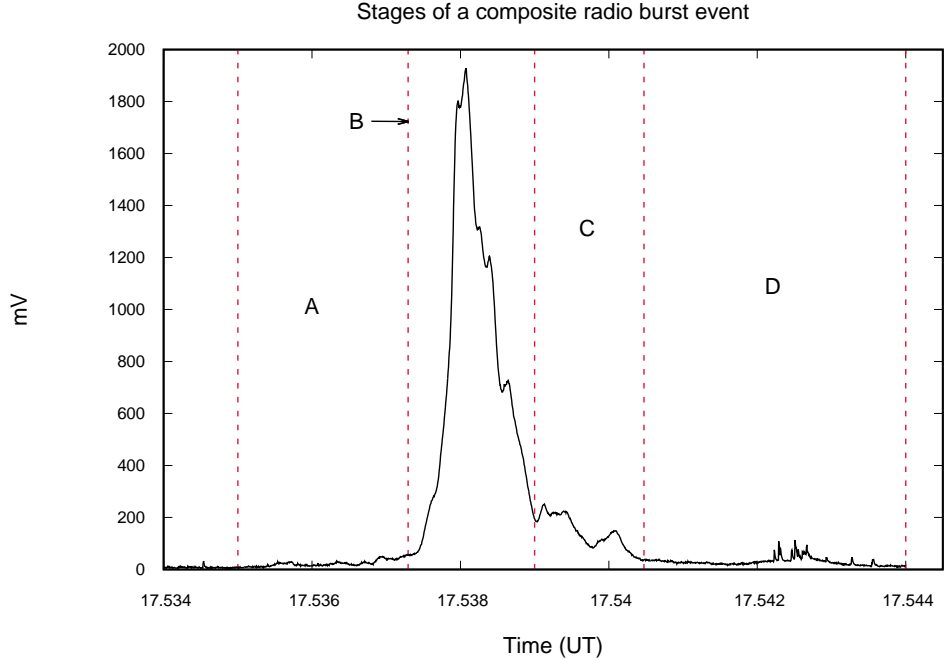}
    \caption{Stages of a single Type III SRB detected by MEXART. This event occurred on 6 September 2017 at 17:32 UT.}
    \label{singleradiob}
\end{figure}

\begin{figure}
    \centering
    \includegraphics[width=1\textwidth]{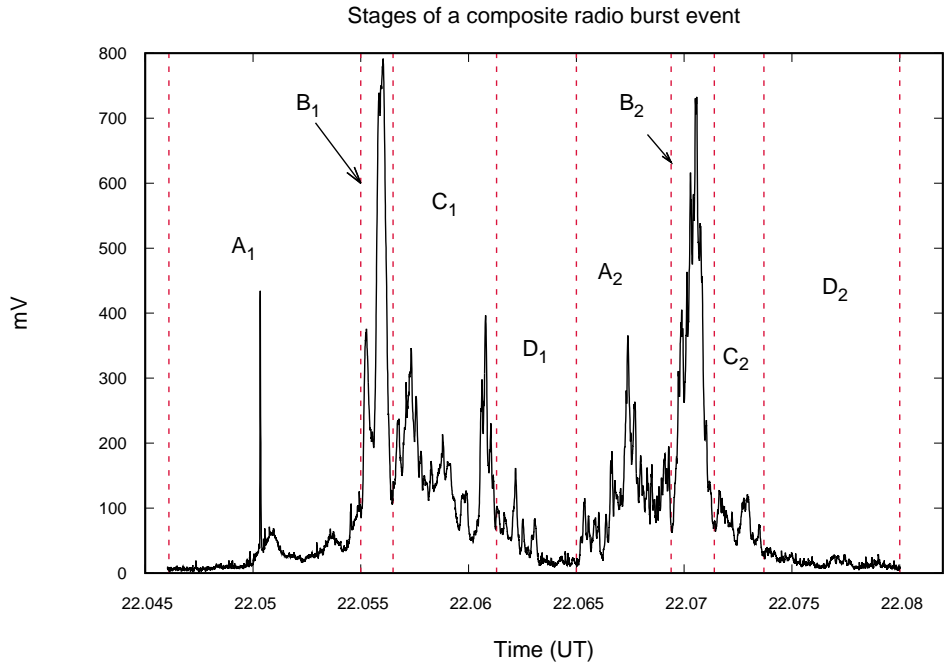}
    \caption{Stages of a composed Type III SRB detected by MEXART. This event occurred on 4 September 2017 at 22:03 UT.}
    \label{composedradiob}
\end{figure}

Representative instances of a single and a composite radio burst structures detected by MEXART are shown in Figures \ref{singleradiob} and \ref{composedradiob}, where the development stages of the radio bursts were labeled with the characters A, B, C, and D: 
A) Pre-phase, a slow increasing signal from the background until the onset of the main stage. B) Main peak, a sudden increase reaching a peak in the radio flux and a sudden decrease forming an overall pulse. C) A decay phase where the signal decreases with an irregular variation with less intensity in the light curve. D) Post-event phase, a slow decrease of the signal reaching the level of  the background.
In several events stages C and D are joined in one single stage (CD). A subscript 1 and 2 in the stages has been used to indicate the corresponding element (single radio burst) of a composed radio burst. 

{}

An analysis of the radio-burst stages has been carried out considering the initial and final boundaries obtained from the light curves of the MEXART observations. The duration of the stages A, B, and CD falls in the range of $0.007-2.664$, $0.011-1.193$, and $0.012-8.730$ minutes respectively. For the three stages most of the durations fall in a interval between $0.00$ and $0.6$ minutes (from which A $\approx 77\,\%$, B $\approx 85\,\%$, and CD $\approx 70\,\%$). 

\begin{figure}
    \centering
    \includegraphics[width=1\textwidth]{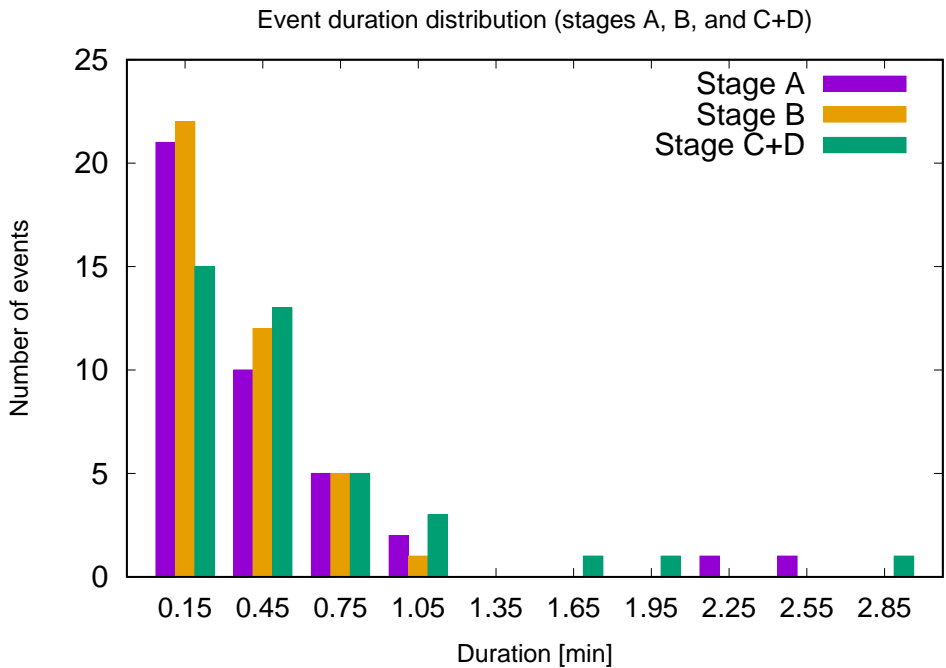}    
    \caption{Distribution of the duration of stages A, B, and CD of the Type III SRB observed by MEXART.}
    \label{histoABCD}
\end{figure}

A histogram with 0.3-minute intervals is shown in Figure  \ref{histoABCD} in order to show the distribution of the radio-burst duration for stages A, B, and CD. From the figure it can be observed that the distribution of the durations exhibits a behaviour that can be fitted with an exponential behaviour of the form $D(t)=b\,e^{kt}$. The values obtained for the parameters $b$ an $k$ that correspond to the stages A, B, and C are shown in  Table \ref{tab:fitparam}, with a negative exponential result for the three cases.

{}

\subsection{Wavelet Representation of SRB Type III}\label{wal}
\begin{figure}
    \centering
    \includegraphics[width=1 \textwidth]{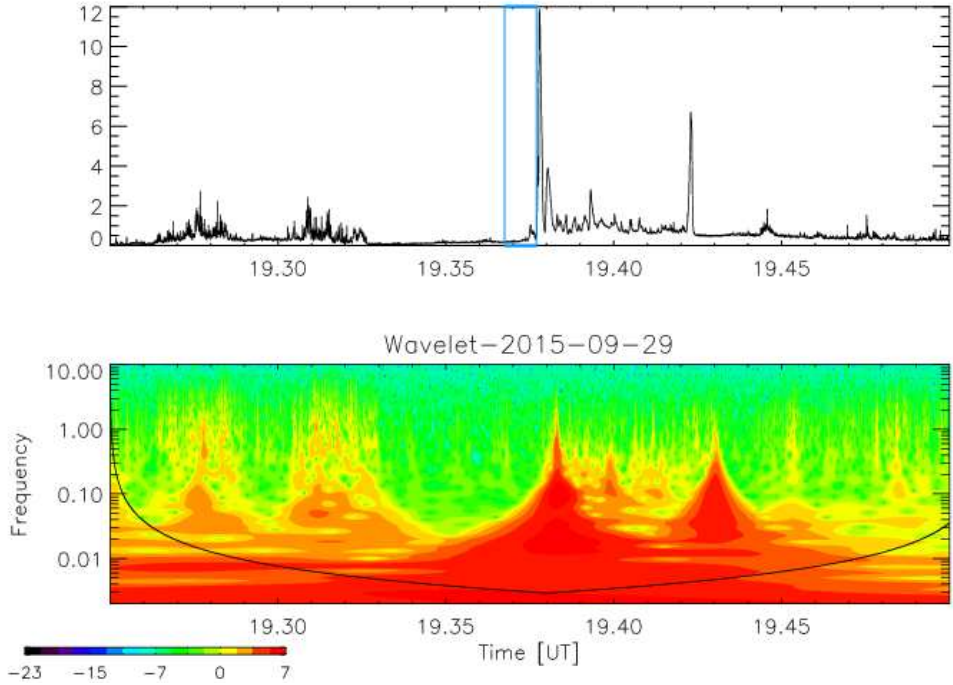}
    \caption{\textit{Above}: A time series observed by MEXART in millivolts
    showing a spike produced by Type III event. A blue rectangle is placed
    at the pre-event. \textit{Below}: A wavelet transform of the observation
    shows an increment in the power at higher frequencies after the
    SRB. The area below the curved line is unreliable as the edge effects of the transform becomes important.}
    \label{wl}
\end{figure}

\begin{figure}
    \centering
    \includegraphics[width=1 \textwidth]{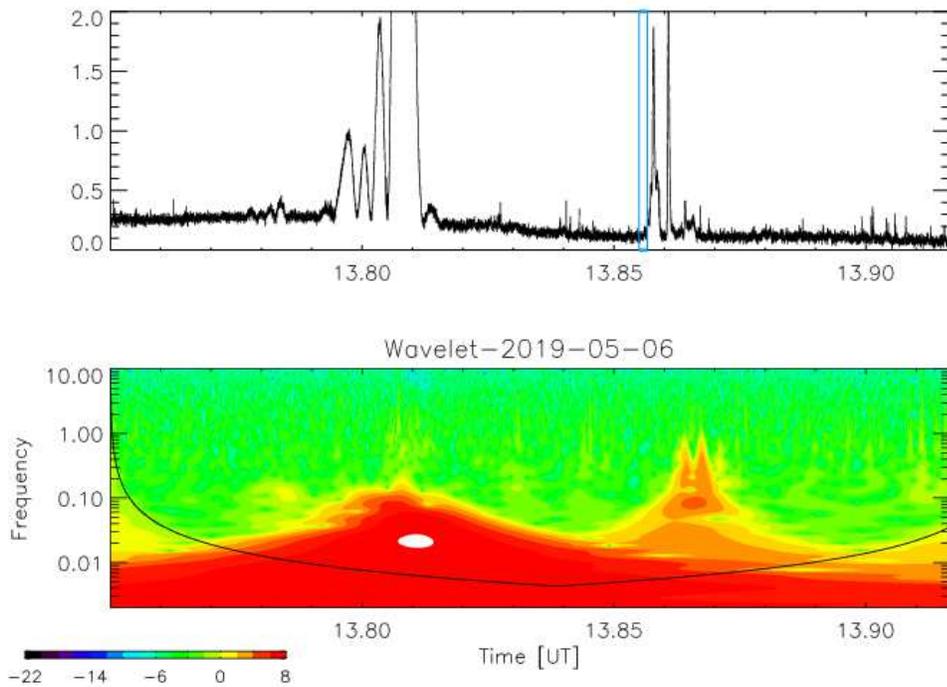}
    \caption{Example of lightcurve (millivolts signal) and wavelet transform of two subsequent signals in the MEXART time series, first at 13.80, a distant radio-source transit, secondly at  13.85, a Type III SRB. The SRB wavelet transform presents its distinctive pine tree shape while the radio-source transit shows a different smoothed shape. The area below the curved line is unreliable as the edge effects of the transform becomes important.}
    \label{nopine}
\end{figure}

By using the wavelet transform, the oscillation frequency of noise fluctuation can be identified for each time of observation, and a power spectrum of the fluctuations can be
displayed for each time. Figure \ref{wl} shows a Morlet wavelet transform applied to a MEXART time series with a Type III SRB, event verified by CALLMEX on 29 September  2015 (see third row of Figure \ref{mosaicII}). The transform shows an increment in noise fluctuations with a pine tree shape following the Type III emission. The blue box in the time series indicates the A stage of the event, meaning the beginning of the event is observable since 19.3675, but with the Morlet wavelet transform, the spectrum indicates the beginning of the event at 19.35, 53 seconds earlier.  The wavelet also shows solar noise storms around minute 19.30, since this observation is near the solar transit during solar activity.
The transform was applied in another 11 Type III events and
this pine tree signature was 
detected 
identified in all of them, so this pine tree
shape is characteristic of the Type III events as other emissions have different spectra. As an example, Figure \ref{nopine} shows two emissions, one corresponding to a radiosource transit and later a Type III SRB; the wavelet transform has a smoothed shape for the first signal and the pine tree shape for the SRB.

\section{Conclusions}\label{conclusions}

To develop this work, the operation and features from both antennas, CALLISTO and MEXART,  were reviewed. Also with the review 
of the different radiosources, many radioemissions, of anthropogenic and natural origin were recorded. Among them we found activity from radiocommunications, electronic devices, electric storms, and solar radio bursts. 
Solar radio bursts were detected and validated in Mexico for  the first time. It was demonstrated that these
bursts are emitted in the communications bands so possible interference effects should be analyzed.

On the other hand, no influence of different radioemissions from astronomical sources occurred during the emission of SRB in MEXART records, according to the \textit{Low-Frequency Array} (LOFAR) \textit{High-Band Array} observations of the Herschel--ATLAS North 
Galactic Pole survey area 
and LOFAR Two-metre Sky Survey DR1 source catalog.





Additionally, some of these records have been related to solar space-weather events. CALLMEX spectra and lightcurves were automatically generated and were archived in the RICE repository \cite{Delaluz2018}. 
The continuous monitoring of events in the 45\,--\,225\,MHz band, in the period of February 2014 to May 2019, allowed the detection of 156  SRB,  from  which 7 are Type II,  one  is  Type I,  and  148  are Type III  from  which 30 are accompanied by a Type V emission.

When relating the events from CALLMEX with MEXART it was found that even when MEXART is focused on IPS studies and has a narrow beam pattern it is sensitive to SRB in an isotropic way. Also e-CALLISTO is fundamental as support to validate SRB in other narrow-band radioastronomy projects.
MEXART identified 74 of these SRB with the aid of CALLMEX.
From these 74, 66 were well recorded at 140\,MHz in CALLMEX, but in another 8 cases, CALLMEX did not recorded them in its 140\,MHz channel, showing that indeed MEXART has a greater sensitivity.

In the MEXART case, from the 156 events, in 5 cases the radiotelescope was not operating and 74 events were not detected, probably due to the  directivity,  the configuration of the instrument, noise of the background, or simply because the SRB did not have had a detectable 140\,MHz emission.

Analyzing the detected Type III SRB within MEXART records, we identified 40 single, clear, well defined events. In these events we identified four phases of the event,  pre-phase (A), main peak (B),  decay phase (C), and post-event (D).

Measuring the durations of each stage and plotting them in a histogram we found a exponential distribution with the form $D(t)  = b\,e^{kt} $ with negative exponential for stages A, B, and CD.

With the MEXART records a Morlet wave transform was performed on 12 Type III events and in their spectra it was found a pine tree structure characteristic of the SRB. Additionally, this structure, seen as a noise fluctuation, is detected in the spectrum before it can be identified in the lightcurve, meaning that the perturbation generating the SRB can be detected before the main emission for 140\,MHz.
Recently, the MEXART radio telescope upgraded and fully digitalized its backend system. The digital system solved the problems with the beamforming and significantly improved the instrument's performance. The MEXART radio telescope now has new technical capabilities: several configuration modes of beams, higher sensitivity, wider bandwidth (12.5\,MHz), 512 frequency channels, and higher sampling rate (1 ms) \cite{gonzalez2021first}. These new technical capabilities of the radio telescope would allow us to perform further studies on solar radio bursts. The new MEXART data could be compared with the CALLMEX observations to infer more information of the four phases of the event in different frequency channels.
The continuous monitoring of the electromagnetic spectrum and the relation of radio events with eruptive solar events 
turns CALLMEX and MEXART into fundamental tools for space weather in Mexico, as it works a the first detector to the country
for space-weather events after they occur and for detailed event analysis.

\section*{Acknowledgments}
Thanks to the Institute of Astronomy, to the ETH Zurich, and to  FHNW Windisch, Switzerland for access to the e-CALLISTO network,  for the stations Birr Castle and Cohoe Alaska data. Thanks for the ROSWELL-NM data who is operated by Stan Nelson.
The authors especially thank Pablo Villanueva and Ernesto Andrade (MEXART staff) for providing valuable information of the observations.
 Victor De la Luz thanks Conacyt Ciencia Basica program 254497 and UNAM PAPIIT TA101920.
 LANCE acknowledges partial support from Conacyt-AEM Grant 2017-01-292684 and LN 315829.
 The authors thanks  PAICyT grant CN913-19.
 This research has made use of SunPy v0.8.3 \cite{stuart_mumford_2018_1160820}, an open-source and free community-developed solar data analysis Python package \cite{2015CS&D....8a4009S}.
 This research made use of Astropy (\url{www.astropy.org}) a community-developed core Python package for Astronomy \cite{astropy:2013, astropy:2018}. 
{\footnotesize\paragraph*{Disclosure of Potential Conflicts of Interest}
The authors declare that they have no conflicts of interest.
}




\begin{table}
\centering
\caption{Parameters of the fitting model $b\,e^{kt}$ for the duration of stages A, B, and CD of the radio burst analyzed.}
\label{tab:fitparam}
\begin{tabular}{l c c c c c}
\hline
 RB  	&    	            & Standard  	    &  	                & Standard       \\
 stage 	&      	  $b$              & error ($b$) $\%$ 	&   	 $k$                & error ($k$) $\%$ \\
\hline
 A      &  $31.4344\pm 1.494$	& 4.751	            &  $-2.54085\pm 0.1586$ & 6.243          \\
 B 	    &  $33.1248\pm 3.139$	& 9.477	            &  $-2.43595\pm 0.3114$ & 12.78	         \\
 CD     &  $21.1094\pm 2.405$	& 11.39	            &  $-1.72318\pm 0.2715$	& 15.76	         \\
\hline
\end{tabular}
\end{table}

\begin{longtable}[c]{llll}
\caption{Table of SRBs detected by CALLMEX in the 140\,MHz channel or by MEXART. The field SRB ID indicates the date and time of the event with the format YYYYMMDDhhmm, the field MEXART indicates if the instrument detected the event, the field CALLMEX 140\,MHz indicates if CALLMEX detected the event at his 140\,MHz channel. Also there are two pairs of SRB IDs repeated, marked with the superscript \textit{a} and \textit{b} which indicates two events, one event Type II and one 
Type III which happened at the same time.}
\label{tableofallevents}\\
\textbf{SRB ID}       & \textbf{Type}  & \textbf{MEXART}  & \textbf{CALLMEX\,140 MHz} \\
\endfirsthead
\multicolumn{4}{c}%
{{\bfseries Table \thetable\ continued from previous page}} \\
\textbf{SRB ID}       & \textbf{Type}  & \textbf{MEXART}  & \textbf{CALLMEX\,140\,MHz} \\
\endhead
201709071343 & I     & no      & yes             \\
201510161328 & II    & yes     & yes             \\
201511041343 & II    & yes     & no              \\
201607100101 & II    & yes     & yes             \\
$201709121914^\textbf{a}$ & II    & yes     & yes             \\
$201905071327^\textbf{b}$ & II    & yes     & yes             \\
201509291922 & III   & yes     & yes             \\
201602112159 & III   & no      & yes             \\
201605021942 & III   & yes     & no              \\
201605251252 & III   & yes     & yes             \\
201703301751 & III   & yes     & yes             \\
201704012138 & III   & no      & yes             \\
201704012144 & III   & no      & yes             \\
201704012301 & III   & yes     & yes             \\
201704012352 & III   & yes     & yes             \\
201704022111 & III   & no      & yes             \\
201704020028 & III   & yes     & yes             \\
201704042134 & III   & yes     & no              \\
201706231924 & III   & no      & yes             \\
201708171822 & III   & yes     & yes             \\
201708181547 & III   & yes     & yes             \\
201709041719 & III   & no      & yes             \\
201709071724 & III   & yes     & yes             \\
201709071729 & III   & yes     & yes             \\
201709071841 & III   & no      & yes             \\
201709072030 & III   & yes     & yes             \\
201709072055 & III   & yes     & yes             \\
201709072202 & III   & yes     & yes             \\
201709081657 & III   & yes     & yes             \\
201709081808 & III   & yes     & yes             \\
201709082012 & III   & yes     & yes             \\
201709082320 & III   & yes     & yes             \\
201709101556 & III   & yes     & yes             \\
201709121531 & III   & yes     & yes             \\
201709251538 & III   & yes     & yes             \\
201709251549 & III   & yes     & yes             \\
201709251711 & III   & yes     & yes             \\
201709251838 & III   & yes     & yes             \\
201709251927 & III   & yes     & yes             \\
201710040013 & III   & yes     & yes             \\
201710052321 & III   & yes     & yes             \\
201710111450 & III   & yes     & yes             \\
201711162040 & III   & no      & yes             \\
201802281610 & III   & yes     & yes             \\
201707112023 & III   & yes     & yes             \\
201707111652 & III   & yes     & yes             \\
$201709121914^\textbf{a}$ & III   & yes     & yes             \\
201802112144 & III   & no data & yes             \\
201802112312 & III   & no data & yes             \\
201803301754 & III   & yes     & yes             \\
201803301800 & III   & yes     & yes             \\
201802112148 & III   & no data & yes             \\
201803302342 & III   & yes     & no              \\
201905061351 & III   & yes     & no              \\
201905061627 & III   & yes     & yes             \\
201905061634 & III   & yes     & yes             \\
201905061644 & III   & yes     & no              \\
201905061656 & III   & yes     & yes             \\
201905061701 & III   & yes     & yes             \\
201905061714 & III   & no      & yes             \\
201905061727 & III   & yes     & yes             \\
201905061748 & III   & yes     & yes             \\
201905061838 & III   & yes     & yes             \\
201905061856 & III   & yes     & no              \\
201905061903 & III   & yes     & yes             \\
201905061905 & III   & yes     & yes             \\
201905061915 & III   & yes     & yes             \\
201905061943 & III   & yes     & yes             \\
201905061949 & III   & yes     & yes             \\
201905061951 & III   & yes     & yes             \\
201905062042 & III   & yes     & yes             \\
201905062214 & III   & yes     & yes             \\
201905062236 & III   & yes     & yes             \\
201905062313 & III   & yes     & yes             \\
$201905071327^\textbf{b}$ & III   & yes     & yes             \\
201905071842 & III   & no      & yes             \\
201605261346 & III,V & yes     & yes             \\
201708170034 & III,V & no      & yes             \\
201708171351 & III,V & no      & yes             \\
201708182103 & III,V & yes     & yes             \\
201709042203 & III,V & yes     & yes             \\
201709061732 & III,V & yes     & yes             \\
201709072314 & III,V & yes     & yes             \\
201709081322 & III,V & yes     & yes             \\
201709081701 & III,V & yes     & yes             \\
201709121543 & III,V & yes     & yes             \\
201905062016 & III,V & yes     & no              \\
201905062137 & III,V & yes     & yes             \\
201905062337 & III,V & yes     & yes             \\
201905062351 & III,V & yes     & yes             \\
201905071334 & III,V & yes     & yes            
\end{longtable}

\begin{longtable}{|c|c|c|}
\caption{Type II SRB related with CME and other events in a 3-hour window after the SRB. The CME records where obtained from \url{kauai.ccmc.gsfc.nasa.gov/DONKI}  . }
\label{typeiicme}\\
\hline
\textbf{SRB }           & \textbf{CME start time}  & \textbf{Linked Events}\\
\hline
\endfirsthead
\multicolumn{3}{c}%
{{\bfseries Table \thetable\ continued from previous page}} \\
\hline
\textbf{SRB }           & \textbf{CME start time}  & \textbf{Linked Events}\\
\hline
\endhead
201710202330 & 20171021T01:18 &                          \\
\hline
201710202330 & 20171021T00:12 & 20171020T23:10 M1,1Flare  \\
\hline
201709121914 & 20170912T20:42 &                          \\
\hline
201511041343 & 20151104T14:24 &  20151104T13:30 M3,7Flare\\
\hline
\end{longtable}

\begin{longtable}{|p{1.1cm}|p{2.7cm}|p{2.7cm}|p{1cm}|p{2.4cm}|}
\caption{The SRB detected by CALLMEX are related by another space-weather events like solar particle events, CME, and flares. For this relation,  the SRB times were compared in a six hour window with the other events times. There are some SRB IDs which are repeated, meaning they reached many other eruptive events. To make distinction of these SRB, a letter is marked at the end of the ID, meaning that all the IDs with the same letter are the same SRB event. These solar eruptive events records where obtained from \url{kauai.ccmc.gsfc.nasa.gov/DONKI}  .}
\label{srbiiicmetable}\\

\hline
\textbf{SRB Count } & \textbf{SRB date }  & \textbf{Flare Peak-Time}    & \textbf{Flare Class}& \textbf{Linked Events} \\
\hline
\endfirsthead
\multicolumn{5}{c}%
{{\bfseries Table \thetable\ continued from previous page}} \\
\hline
\textbf{SRB Count } & \textbf{SRB date }  & \textbf{Flare Peak-Time}    & \textbf{Flare Class}& \textbf{Linked Events} \\
\hline
\endhead
7         & 201704011931, 201704012114\textbf{a}, 201704012138\textbf{b}, 201704012144\textbf{c}, 201704012301, 201704012352, 201704020028\textbf{d} & 2017-04-01 T21:48 & M4.4       & 2017-04-01 T22:12CME                                                                                                     \\\hline
6         & 201704011703, 201704011712, 201704011931, 201704012114\textbf{a}, 201704012138\textbf{b}, 201704012144\textbf{c}                 & 2017-04-01 T19:56 & C3.7       & 2017-04-01 T20:12CME                                                                                                     \\\hline
5         & 201709072119, 201709072151, 201709072202, 201709072233, 201709072314                                 & 2017-09-07 T23:59 & M3.9       & 2017-09-07 T23:36CME                                                                                                     \\\hline
4         & 201709081657, 201709081808, 201709081322, 201709081701                                                 & 2017-09-08 T15:47 & M3.0       &                                                                                                                            \\\hline
2         & 201709041719\textbf{e}, 201709042203\textbf{f}                                                                                 & 2017-09-04 T19:37 & M1.7       &                                                                                                                            \\\hline
2         & 201709071724, 201709071729                                                                                 & 2017-09-07 T14:36 & X1.3       & 2017-09-07 T15:12CME                                                                                                     \\\hline
2         & 201704012352, 201704020028\textbf{d}                                                                                 & 2017-04-02 T02:46 & C8.0       &                                                                                                                            \\\hline
2         & 201709041719\textbf{e}, 201709042203\textbf{f}                                                                                 & 2017-09-04 T20:02 & M1.5       &                                                                                                                            \\\hline
1         & 201709041719\textbf{e}                                                                                                 & 2017-09-04 T18:22 & M1.0       & 2017-09-04 T19:39CME                                                                                                     \\\hline
1         & 201709101556                                                                                                 & 2017-09-10 T16:06 & X8.2       & 2017-09-10 T16:09CME,  2017-09-10 T16:25SEP,  2017-09-10 T16:45SEP,  2017-09-10 T20:00SEP                          \\\hline
1         & 201602112159                                                                                                 & 2016-02-11 T21:03 & C8.9       & 2016-02-11 T21:28CME                                                                                                     \\\hline
1         & 201709042203\textbf{f}                                                                                                 & 2017-09-04 T20:33 & M5.5       & 2017-09-04 T20:36CME,  2017-09-04 T22:56SEP,  2017-09-04 T23:52SEP,  2017-09-05 T00:30SEP,  2017-09-06 T23:45SEP \\\hline
1         & 201709042203\textbf{f}                                                                                                 & 2017-09-04 T22:14 & M2.1       &                                                                                                                            \\\hline
1         & 201709041719\textbf{e}                                                                                                 & 2017-09-04 T15:30 & M1.5       & 2017-09-04 T19:39CME                                                                                                     \\\hline
1         & 201709061732\textbf{g}                                                                                                 & 2017-09-06 T19:30 & M1.4       &                                                                                                                            \\\hline
1         & 201604180024                                                                                                 & 2016-04-18 T00:29 & M6.7       & 2016-04-18 T00:36CME                                                                                                     \\\hline
1         & 201704022111\textbf{h}                                                                                                 & 2017-04-02 T20:33 & M5.7       &                                                                                                                            \\\hline
1         & 201704022111\textbf{h}                                                                                                 & 2017-04-02 T18:38 & M2.1       & 2017-04-02 T20:09CME                                                                                                     \\\hline
1         & 201709082320                                                                                                 & 2017-09-08 T23:45 & M2.1       &                                                                                                                            \\\hline
1         & 201709061732\textbf{g}                                                                                                 & 2017-09-06 T15:56 & M2.5       &                                                                                                                            \\\hline
1         & 201802112312                                                                                                 & 2018-02-12 T01:35 & C1.5       & 2018-02-12 T01:25CME                                                                                                     \\\hline
1         & 201709121914                                                                                                 & 2017-09-12 T19:20 & C1.6       & 2017-09-13 T05:33SEP                                                                                                     \\\hline
1         & 201509291922                                                                                                 & 2015-09-29 T19:24 & M1.1       &                                           \\\hline                                                                                
\end{longtable}

\bibliography{huipe-domratcheva-2019-rev1} 
\bibliographystyle{unsrtnat}
\end{document}